\documentclass[9pt,onecolumn,twoside]{arxiv}
\usepackage[]{geometry}
\usepackage[scriptsize,tight]{subfigure}
\usepackage{graphicx}
\usepackage{cite}
\usepackage{multirow}
\usepackage{color,soul}
\usepackage{hyperref}

\articletype{inv} 

\runningtitle{Adsorption of Ionized Glyphosate on COOH--Modified Carbon Nanotubes}

\runningauthor{Silva \textit{et al.}}

\title{From Electronic Structure to Environmental Remediation: Adsorption of Ionized Glyphosate on COOH--Modified Carbon Nanotubes}

\author[1]{H.~T.~Silva}
\author[1]{L.~C.~S.~Faria}
\author[1]{C.~Aguiar}
\author[1]{T.~A.~Aversi-Ferreira}
\author[1,$\ast$]{I.~Camps}
\correspondingauthoraffiliation[$\ast$]{icamps@unifal-mg.edu.br}

\affil[1]{Laborat\'orio de Modelagem Computacional - \emph{La}Model,
	Instituto de Ci\^{e}ncias Exatas - ICEx. Universidade Federal de Alfenas -
	UNIFAL-MG, Alfenas, Minas Gerais, Brazil}

\begin{abstract}
	Glyphosate is a widely used herbicide whose persistence and toxicity in aquatic and terrestrial environments demand efficient removal strategies. Here we employ GFN2 xTB calculations with implicit ALPB aqueous solvation and automated docking to investigate the adsorption of all five pH dependent ionized forms of glyphosate (G1--G5) on (10,0) single walled carbon nanotubes covalently functionalized with carboxyl groups at 0--25\% coverage. Adsorption energies reveal a clear charge dependent trend: the dianionic (G4) and trianionic (G5) species exhibit the most negative binding energies over the entire functionalization range, while the protonated and neutral forms (G1, G2) bind weakly, approaching reversible adsorption at high COOH contents. The deprotonated form G5 strengthens from -2.17~eV on pristine CNT to about -5.7 eV at 10\% and 25\% functionalization, with a local minimum near 15--20\% COOH due to steric and electrostatic crowding of adjacent groups. Decomposition of the solvation free energy shows dominant electrostatic stabilization complemented by increasingly favorable hydrogen bond contributions as carboxyl density grows. All complexes display very small HOMO--LUMO gaps (0.02--0.22~eV), indicating high electronic sensitivity to adsorption and functionalization. Overall, CNT+COOH systems can operate in both strong capture and regenerable regimes depending on glyphosate ionization state, offering a tunable platform for pH responsive remediation.
\end{abstract}


\begin{document}
\maketitle
\thispagestyle{firststyle}
\vspace{-13pt}

\section{INTRODUCTION}
\label{Sec:Intro}
The intensive use of pesticides and herbicides in modern agriculture represents one of the major paradoxes of modern societys needs and development, while ensuring large-scale productivity, it has progressively compromised the integrity of ecosystems and the health of exposed populations~\cite{Shekhar-ToxicolRep-13-101840-2024,Munoz-Bautista-Agronomy-15-1878-2025}. Herbicides are widely used in agriculture due to their efficiency, protecting crops from pests and ensuring a favorable cost-benefit ratio to improve the quality and yield of agricultural production, thereby maintaining global food security~\cite{Zhou-EmergingContaminants-11-100410-2025,Lazarevic-Pasti-Foods-14-1128-2025}. In this context, glyphosate (GLY), a chemical compound derived from glycine, is a broad-spectrum herbicide belonging to the class of phosphonated amino acids, representing one of the most produced and consumed agrochemicals worldwide~\cite{Galli-FrontiersinToxicology-6-1474792-2024,Badani-EurJEnvironSci-13-5-2023}. Its efficiency in weed control, combined with its ease of application, has consolidated its use in agricultural crops, forested areas, and urban zones across virtually all continents~\cite{Lazarevic-Pasti-Foods-14-1128-2025,KimbiYaah-EnvironRes-240-117477-2024}. However, despite being considered safe, theoretical and experimental evidence indicates that pesticide residues can cause severe environmental and toxicological impacts~\cite{Chavez-Reyes-JXenobiot-14-604-2024}. This includes the contamination of soil and water resources, adverse effects on wildlife, and potential human health alterations, such as endocrine disruption and neurological impairments~\cite{Badani-EurJEnvironSci-13-5-2023,Yubolphan-EnvironToxicolPhar-125-105094-2026}. Furthermore, certain substances are classified by the World Health Organization (WHO) as carcinogenic to humans~\cite{Shekhar-ToxicolRep-13-101840-2024,Kalyabina-ToxicolRep-8-1179-2021,Ray-EnvironAnalHealToxicol-38-2023017-2023,KimbiYaah-EnvironRes-240-117477-2024,Barroso-Sustainability-17-3891-2025}.

And although it is not mentioned in the WHO drinking water guidelines, some countries have banned its use in agricultural production, while others have established maximum acceptable concentrations in tap water~\cite{Shekhar-ToxicolRep-13-101840-2024,KimbiYaah-EnvironRes-240-117477-2024}. This has prompted different researchers worldwide to test various technologies to remove or degrade GLY~\cite{Li-EnvironSciPollutR-25-21036-2018,Sittiwong-MicroporMesoporMat-341-112083-2022}. However, there are numerous challenges in scaling up purification technologies due to high costs and a lack of concrete information regarding their adverse effects~\cite{Wang-Chemosphere-331-138827-2023,KimbiYaah-EnvironRes-240-117477-2024}. Further exacerbating this situation, the main degradation product of glyphosate, aminomethylphosphonic acid (AMPA), can be adsorbed onto soil particles and remain retained within the vadose zone~\cite{Singh-IntJEnvResPubHe-17-7519-2020}. Furthermore, studies have shown that exposure to AMPA may cause damage to human erythrocytes and induce chromosomal aberrations in fish, highlighting its potential toxic effects on different organisms~\cite{Grandcoin-WaterRes-117-187-2017,Singh-IntJEnvResPubHe-17-7519-2020,SalgadoKiefer-EnvironToxicolPhar-107-104429-2024}. Given such, the development of functional materials capable of efficiently detecting, adsorbing, and removing glyphosate from different environmental matrices becomes an urgent scientific and technological necessity.

Among the nanomaterials applied to environmental remediation, carbon nanotubes (CNTs) stand out for their high surface area, electrical conductivity, and mechanical resistance~\cite{Saleh-JIndEngChem-146-176-2025}. These properties favor the adsorption of organic pollutants through $\pi-\pi$ interactions, hydrogen bonding, van der Waals forces, and electrostatic interactions, making them promising for the removal of contaminants such as dyes, pharmaceuticals, and pesticides~\cite{Sandoval-ChemRev-126-2283-2026}. However, their hydrophobic nature limits dispersion in aqueous media and interaction with polar contaminants~\cite{Alfei-JXenobiotics-15-76-2025}, making surface modification necessary to expand their applicability in environmental systems. Thus, covalent functionalization is a central strategy to modify the surface of carbon nanotubes (CNTs), reducing their hydrophobicity, increasing hydrophilicity, and elevating reactivity toward target molecules~\cite{Alosta-MaterTodayCommun-50-114504-2026}.

In this context, different oxygen-containing functional groups have been employed to enhance adsorptive performance~\cite{Dong-ChemEngJ-499-156654-2024}. Studies involving functionalization with hydroxyl groups (--OH) have demonstrated an increase in surface hydrophilicity, favoring donor--acceptor interactions and the formation of hydrogen bonds with polar molecules~\cite{Milowska-JChemPhys-138-194704-2013,Rezazade-BMCChemistry-18-85-2024,Silva-ApplSurfSci-729-166060-2026}. In addition, computational simulations indicate that an increase in the concentration of --OH groups intensifies glyphosate adsorption, promoting more favorable binding energies, greater charge transfer, and lower molecular mobility, without compromising the regeneration of the adsorbent in systems with moderate interactions~\cite{Dong-ChemEngJ-499-156654-2024,Silva-ApplSurfSci-729-166060-2026}.

With this in mind, another functionalization strategy would be the introduction of carboxyl groups (--COOH) to increase the adsorptive performance of CNTs~\cite{Mananghaya-JMolLiq-212-592-2015,Dong-ChemEngJ-499-156654-2024,Rezazade-BMCChemistry-18-85-2024}. The incorporation of this group promotes the partial transition from sp$^2$ to sp$^3$ to hybridization at the functionalized sites, raises the surface charge density, creates new proton donor and acceptor sites, and favors the formation of hydrogen bonds with polar molecules~\cite{Veloso-ChemPhysLett-430-71-2006,Lara-ChemPhys-428-117-2014}. In addition to surface functionalization, computational studies have also investigated the interactions between glyphosate in different ionization states and single-walled carbon nanotubes, evidencing the potential of these materials for the detection and capture of the herbicide~\cite{Silva-SurfacesandInterfaces-93-109439-2026}. The results indicate that carbon nanotubes exhibit a higher affinity for the ionized forms of glyphosate (G1, G3, G4, and G5), whereas the neutral form (G2) displays low adsorption. Meanwhile, the fully deprotonated (trianionic) form (G5) presents high adsorption stability, which may hinder the regeneration of the material~\cite{Silva-SurfacesandInterfaces-93-109439-2026}. Given these results, investigating the functionalization of CNTs with carboxyl groups (CNT+COOH) proves relevant, aiming to evaluate their potential in optimizing the adsorption of the different ionized forms of glyphosate.

However, the concentration of the functionalizing agent on the CNTs must be controlled, as insufficient levels can limit adsorption, while elevated levels can compromise their structural, electronic, and mechanical properties~\cite{Bulla-JEnvironChemEng-12-114504-2024}. Furthermore, glyphosate adsorption is strongly influenced by pH, due to its four pKa values (2.0; 2.6; 5.6 and 10.6), which determine different protonation states and, consequently, distinct molecular properties and interactions~\cite{Silva-ApplSurfSci-729-166060-2026}. Thus, considering the variability of pH in natural environments, the analysis of the interaction between CNT+COOH and the five ionized forms of glyphosate (CNT+COOH/glyphosate) allows for a more realistic and environmentally relevant description of the adsorption process.

\section{MATERIALS AND METHODS}
\label{Sec:Method}
We employed single-walled carbon nanotubes of chirality (10,0), a semiconducting zigzag structure with a diameter of 7.83~\AA~and a length of 12.78~\AA. Both historical and practical arguments motivated this choice. Historically, the diameter distributions reported in Iijima's pioneering study~\cite{iijima-Nature-363-603} peaked around 8~\AA~and 10.5~\AA, values that correspond to the (10,0) and (13,0) nanotubes, respectively---a finding that was attributed to helicity and to the underlying growth mechanisms. Practically, (n,0) zigzag nanotubes are achiral (chiral angle of 0\textsuperscript{o}) and therefore possess high azimuthal symmetry, while the (10,0) tube in particular is semiconducting. Its electronic properties are thus easily tuned and verified by means of adsorption or surface functionalization.

The strong C--C interactions that arise from the sp$^{2}$ hybridization of carbon nanotubes considerably reduce their surface activity, so that pristine nanotubes perform poorly as gas filters or sensors for metal ions and organic species. To overcome this limitation and open the surface to coupling with a variety of functional groups, the nanotube is typically subjected to chemical treatments. Grafting groups such as \text{--COOH} and \text{--OH} proves especially effective, since it not only breaks the surface symmetry but also promotes hydrogen bonding and electrostatic interactions. Therefore, both pristine and carboxyl-functionalized forms, at functionalization degrees of 5, 10, 15, 20, and 25\% were used in this work. For each concentration, the representative structure was chosen on the basis of the highest system entropy, using quasi-entropy~\cite{Oganov2009} as the selection criterion~\cite{Ribeiro2017}. 

Glyphosate, on the other hand, is a polyprotic species whose ionization state depends strongly on the pH of the surrounding medium~\cite{Gill_2017,Singh_2024,Evalen_2024}, so that every ionic form displays its own charge distribution and reactivity. Such differences have a direct bearing on the adsorption selectivity and on the binding mechanisms operating at carbon-based materials.

The interactions between glyphosate and functionalized CNTs were investigated with the semi-empirical tight-binding approach available in the xTB (extended tight-binding) package, a self-consistent and accurate framework that accounts for electrostatic multipole contributions as well as for density-dependent dispersion~\cite{Bannwarth_2020}. Molecular geometries were optimized so as to locate the configuration of lowest potential energy, by adjusting the atomic positions. Atomic coordinates were varied systematically until the computed energy attained its minimum, corresponding to the ground state~\cite{Schlegel_2011}. Structures were optimized at an extreme level, adopting an energy convergence criterion of $5\times10^{-8}\,E_h$ together with a gradient-norm convergence of $5\times10^{-5}\,E_h/a_0$, in which $a_0$ denotes the Bohr radius~\cite{Aguiar_2024}.

\begin{figure}[tbph]
	\centering
	\includegraphics[width=15cm]{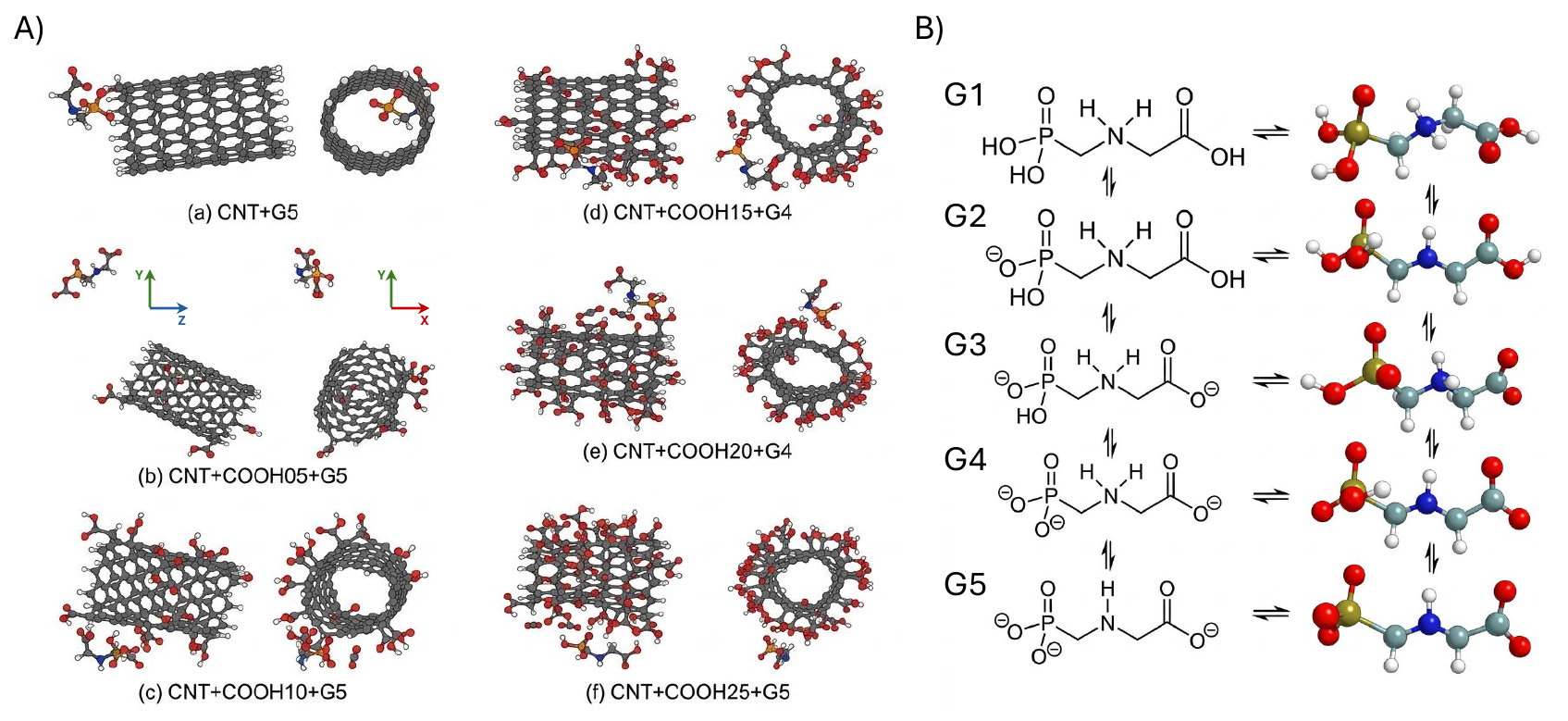}
	\caption{\label{Fig:STRUCALL} Panel A: Structures of the optimized complexes. Panel B: 2D and 3D representations of the different ionized forms of glyphosate as a function of pKa values.  Images of molecular structures rendered with Jmol software~\cite{jmol} using the internal CPK color scheme: gray for carbon, red for oxygen, white for hydrogen, blue for nitrogen and golden for phosphorous. (For interpretation of the references to color in this figure legend, the reader is referred to the web version of this article.)}
\end{figure}

Each complex under study was built from a single-walled carbon nanotube (CNT) functionalized with carboxyl functional group (--COOHx), where x represents concentrations of 0, 5, 10, 15, 20, and 25\% (see Figure~\ref{Fig:STRUCALL}A) and glyphosate taken at a different degree of ionization (see Figure~\ref{Fig:STRUCALL}B). These degrees of ionization follow from the acid dissociation constants of glyphosate (pKa: 2.0, 2.6, 5.6, and 10.6), which in turn are dictated by the pH of the medium hosting the molecule, giving rise to G1 (pH~$< 2$), G2 (pH~$\approx 2-3$), G3 (pH~$\approx 4-6$), G4 (pH~$ \approx 7-10$), and G5 (pH~$> 10.6$)~\cite{Herath_2019}. Systems were labeled as CNT+Gy and CNT+COOHx+Gy, in which CNT stands for carbon nanotube, COOHx for the carboxyl functional group, and Gy for the particular ionized form of glyphosate present in that complex. Protonation states, net charges, and the dominant ionic form for each pH interval considered in this work are collected in Table~\ref{Tab:GlyProto}.

\begin{table}[tbph]
\caption{Glyphosate protonation states, net charges, and domain ionic form.}
\label{Tab:GlyProto}
\begin{center}
\setlength\extrarowheight{-3pt}
\begin{tabular}{crcll}
\hline
System & Net Charge & pH regimes & Protonation states of & Domain ionic \\
              &                      &                       & functional groups       & form         \\
\hline
\hline \\
G1 & +1 & $<$ 2      & NH$^{+}_{3}$, COOH, PO$_{3}$H$_{2}$      & Protonated    \\

G2 &  0 & 2-3     & NH$^{+}_{3}$, COO$^{-}$, PO$_{3}$H$_{2}$ & Zwitterionic  \\

G3 & -1 & 4-6     & NH$^{+}_{3}$, COO$^{-}$, PO$_{3}$H$^{-}$  & Monoanionic    \\

G4 & -2 & 7-10    & NH$^{+}_{3}$, COO$^{-}$, PO$_{3}^{2-}$   & Dianionic      \\

G5 & -3 & $>$10.6   & NH$_{2}$, COO$^{-}$, PO$_{3}^{2-}$      & Deprotonated    \\
\hline
\end{tabular}
\begin{flushleft}
\end{flushleft}
\end{center}
\end{table}

\begin{figure}[tbph]
\centering
\includegraphics[width=13cm]{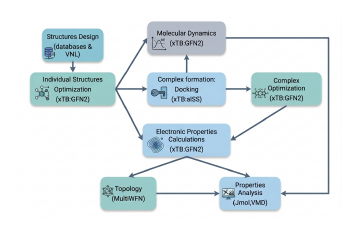}
\caption{\label{Fig:Methods} Flowchart of the computational procedure used for analysis of interactions between carbon nanotube and glyphosate.}
\end{figure}

Calculations followed the sequence of steps laid out in the flowchart of Figure~\ref{Fig:Methods} and were performed with the xTB package (GFN2-xTB) using the ALPB (analytical
linearized Poisson--Boltzmann) implicit solvation model with water as solvent. The
ALPB model provides a self-consistent continuum treatment of aqueous solvation that has been
specifically parameterized for modern xTB methods, enabling a realistic description of
electrostatic screening and hydrogen--bonding stabilization for highly charged, polar species
such as glyphosate and oxidized CNT surfaces at a computational cost compatible with
extensive sampling~\cite{xTB_ALPB}.

 The geometries of the isolated nanotube and of isolated glyphosate were optimized first. The coupling stage came next, handled by automated interaction site mapping (aISS), with the initial arrangement of the glyphosate molecules relative to the CNT generated automatically through the xTB molecular docking protocol. The docking module proceeds as described below. A search for pockets in molecule A is performed first, followed by a screening for $\pi-\pi$ stacking interactions along different directions in three dimensions (3D). Global orientations of molecule B (glyphosate, in the present case) are then explored on an angular grid surrounding molecule A (here, the CNT). Ranking of the resulting structures relies on the interaction energy (xTB-IFF). By default, the 100 structures of lowest interaction energy move on to a subsequent two-step genetic algorithm optimization, which guarantees that conformations missed during the first screening are also taken into account. This two-step genetic optimization consists of a random crossover between each pair of positions of molecule B around molecule A, followed by a random mutation of 50\% of the structures in both position and angle. The entire search procedure is repeated 10 times, and the 10 complexes of lowest interaction energy are retained. The single structure with the lowest interaction energy overall is finally chosen for optimization of the complex and used as input for the molecular dynamics~\cite{xTB-dock}.

Classification of the generated structures relied on the interaction energy, for which the default choice is a two-step refinement protocol built on genetic algorithms: the one hundred structures displaying the lowest interaction energies are retained so that conformations overlooked in the initial screening are still included. Throughout this two-step genetic optimization, the position of each glyphosate molecule was randomly recombined around the carbon nanotube, after which 50\% of the structures were subjected to random mutations affecting both position and angle. Ten iterations of this search process left ten complexes with the lowest interaction energies.

Electronic properties were obtained within the spin polarization scheme, the following quantities being computed for the CNT+COOHx+GY systems: the highest occupied molecular orbital energy (HOMO, $\varepsilon_H$), the lowest
unoccupied molecular orbital energy (LUMO, $\varepsilon_L$), and the gap separating the HOMO and LUMO orbitals ($\Delta \varepsilon = \varepsilon_H - \varepsilon_L$), all evaluated with Fermi-Dirac smearing at an electronic temperature of 300~K.

In the GFNn-xTB methods the electronic temperature is an adjustable parameter. Fermi-Dirac smearing is employed so that fractional orbital occupations become possible, an option that proves particularly valuable for systems with nearly degenerate levels or narrow HOMO--LUMO gaps. Obtaining integer 2/0 HOMO/LUMO occupations would demand lowering the temperature to 0~K, which reduces the Fermi distribution to a step function; such a choice is nonetheless discouraged in GFN methods, since it frequently provokes self--consistent--field (SCF) convergence failures, above all in cases of very small gaps where several states make sizable contributions to static correlation. Far from being a shortcoming, the fractional occupations that emerge constitute a deliberate feature offering an inexpensive approximation to that correlation. Sensitivity of the results to the default electronic temperature of 300~K is modest: for geometry optimizations and for properties evaluated near minima the impact of variations is negligible, whereas in high-temperature molecular dynamics or in electronic structure analyses higher temperatures may affect averages such as bond distances or entropies, benchmarks indicating that these temperature-dependent effects scale approximately linearly up to $\sim$~350~K and grow faster beyond that point~\cite{xTB_GFN2,Mewes_2021}.

The mobility of the charge carriers and the electronic interactions between the nanotube and the glyphosate molecules were probed by computing the atomic charge distribution (within the CM5 scheme~\cite{charges_CM5}) along with the electronic transfer integrals~\cite{Kohn_2023}, both available in the xTB package.

Charge transfer between the isolated CNT and the CNT extracted from the complexes was obtained from equation~\ref{Eq:DeltaQ} below:
\begin{equation}
\label{Eq:DeltaQ}
\Delta Q = Q^{ads}_{CNT} - Q^{iso}_{CNT},
\end{equation}
where $Q^{ads}_{CNT}$ is the total charge of the CNT after adsorption and $Q^{iso}_{CNT}$ is the total charge for the isolated CNT.


The adsorption energy ($E_{ads}$), defined as the difference between the energy of the final CNT+Gy system ($E_{CNT+Gy}$) and the sum of the energies of the initial isolated CNT ($E_{CNT}$) and glyphosate ($E_{Gy}$) systems, reads:

\begin{equation}
\label{Eq:bind}
E_{ads} = E_{CNT+Gy} - E_{CNT} - E_{Gy}.
\end{equation}

The counterpoise correction, which compensates for the effects of the basis set superposition error (BSSE), is customarily applied when interaction energies of small molecules are computed, so as to prevent these energies from being under- or overestimated~\cite{bsse}. Since xTB relies on a minimal, atom-centered basis, it remains formally exposed to basis-set superposition errors, exactly as any other finite-basis quantum method. In practice, though, BSSE in methods of the GFN-xTB type is to a large extent absorbed by the empirical parametrization and by the accompanying corrections~\cite{xTB_GFN2}, which is why it is not usually addressed explicitly (through a counterpoise correction, for instance) and why it tends to be smaller than in low-cost \emph{ab initio} calculations employing small Gaussian basis sets. Where noncovalent interaction energies are concerned, the leading uncertainties in xTB normally stem from the underlying semiempirical approximations and from the dispersion model rather than from ``classical'' BSSE, even if some caution regarding overbinding is still advisable, together with verification of the key trends against higher-level data whenever that is feasible.

Aiming to understand and classify the nature and strength of the interactions established between glyph\-o\-sate and the carbon nanotube, a study of topological properties was carried out on the basis of the wave function delivered by the calculation of electronic properties~\cite{Aguiar_2024}. Bond critical points (BCPs) could thereby be identified and descriptors such as the electron density ($\rho$), the Laplacian ($\nabla^2 \rho$), the electronic localization function (ELF), the localized orbital locator (LOL), the local kinetic (G(r)), potential (V(r)), and total (H(r)) energy densities; their analysis was performed with the MULTIWFN software, which takes as input the wave function generated during the calculations of electronic properties~\cite{Lu-JComputChem-33-580-2012,Multiwfn2}. In assessing the strength and the type of bond between attractive pairs of atoms, attention was restricted to bond critical points of type \textbf{(3,-1)}, since these are marked by a minimum of the electron density along the bond path connecting two nuclei at the interface between the glyphosate molecule and the nanotube. The parameters so obtained rest on a physically observable quantity (the electron density) and are free of bias, thus complementing the techniques based on the wave function or on molecular orbital analysis. Physical meaning is in this way not attributed to one particular set of orbitals, for analyses supported exclusively on them, while not devoid of physical content, may miss relevant details. The electron density carries the further advantage of being amenable to both theoretical and experimental scrutiny~\cite{Bader1994,Koch_2024,Fedorov_2025}.

Whereas geometry optimization aims at locating the lowest energy structure on a potential energy surface, molecular dynamics (MD) simulations gave us access to the motion of the glyphosate molecules, thus affording a fuller picture of the dynamic behavior of the system~\cite{Martinez_2003}. These simulations were run at $300\,K$ over a production time of 100~ps, adopting a time step of 2~fs and a dump step of 50~fs, the final configuration being stored in a trajectory file. The GFN-FF force field was used in such calculations, having been designed precisely to reconcile high computational efficiency with the accuracy normally ascribed to quantum mechanical methods~\cite{xTB_GFN-FF}.

Temperature control in the molecular dynamics module of the xTB software is exercised by the Berendsen~\cite{Berendsen_2007} thermostat during NVT ensemble simulations~\cite{xTB_MTD}. Velocities are gently rescaled by this thermostat in order to keep the target temperature, which yields stable temperature regulation without, however, producing a genuine canonical ensemble. Its simplicity and efficiency render the Berendsen thermostat well suited to fast semiempirical MD simulations of the kind performed in xTB, though its weak coupling behavior, when set against other thermostats, may leave an imprint on dynamic properties.

The spatial distribution of the glyphosate molecules was characterized through the radial distribution function (RDF):

\begin{equation}
\label{Eq:RDF}
g(\bf{r}) = \frac{n(\bf{r})}{4 \pi \rho \bf{r}^2 \Delta \bf{r}},
\end{equation}
where $n(\bf{r})$ is the mean number of particles in a shell of width $\Delta \bf{r}$ at distance $\bf{r}$, and $\rho$ is the mean particle density.

Statistically, $g(\bf{r})$ describes the probability of finding a glyphosate molecule at a position $\bf{r}$ relative to the carbon nanotube, normalized by the average density. The analyses considered all atoms of the glyphosate molecule and all atoms of the carbon nanotube (CNT), without invoking center-of-mass approximations or filtering by specific atom type. Atomic selections were defined according to the residue assignments of each molecule so as to ensure a comprehensive evaluation of the spatial distribution in the interfacial regions. This approach is well suited to heterogeneous systems such as the one studied here, as it helps to predict how glyphosate organizes itself with respect to the nanotube surface~\cite{Hansen_2013}.

\section{RESULTS AND DISCUSSION}
\label{Sec:Results}

\subsection{Structure and Electronic Analysis}
\label{ElecStruc}
The optimized complexes resulting from geometry optimization are shown in Figure~\ref{Fig:STRUCALL}A in both lateral (left) and frontal (right) views, corresponding to projections along the $x$-axis and $z$-axis, respectively. As previously reported~\cite{Ribeiro2017}, radial deformation of the nanotubes increases with functionalization concentration. To characterize the structural deformations induced in the nanotubes by functionalization and by the interaction with glyphosate, the root-mean-square deviation (RMSD) of the atomic positions was computed between two optimized structures: the pristine nanotube (zero functionalization concentration, no glyphosate) and the ``bare'' nanotubes, that is, structures from which the functional groups and/or glyphosate molecules had been manually removed. The $\Delta$RMSD is then evaluated between the backbones (carbon atoms of the nanotube only) of the functionalized systems without and with glyphosate, for instance, between the CNT+COOH05 and CNT+COOH05+G5 systems. The calculated $\Delta$RMSD values for all systems are collected in Table~\ref{Tab:ElectRes}. Positive values indicate that interaction with glyphosate increases the backbone deformation, whereas negative values indicate that glyphosate contributes to structural stabilization; in other words, $\Delta\text{RMSD} > 0$ means glyphosate distorts the nanotube backbone, while $\Delta\text{RMSD} < 0$ means it promotes a more ordered structure.

\begin{table}[tbph]
	\caption{Results from structural and electronic calculations$^\dag$.}
	\setlength{\belowcaptionskip}{-30pt} 
	\label{Tab:ElectRes}
	\begin{center}
		\setlength\extrarowheight{-3pt}
		\begin{tabular}{lrlllllll}
			\hline
			System & $\Delta$RMSD                                                   & $E_{ads}$ & $\varepsilon_H$ & $\varepsilon_L$ & $\Delta \varepsilon$ & $\Delta G_{Elec}$ & $\Delta G_{SASA}$ & $\Delta G_{HB}$ \\ \hline \hline
			CNT+G5   &   0.0                                              & -2.17     & -8.42           & -8.39           & 0.04                 & -14.35            & -0.21             & -0.13           \\
			CNT+COOH05+G5 & 8.6 & -3.13     & -8.45           & -8.23           & 0.22                 & -12.03            & 0.34              & -1.75           \\
			CNT+COOH10+G5 & -17.0                                             & -5.69     & -8.29           & -8.22           & 0.07                 & -13.94            & 0.58              & -2.91           \\
			CNT+COOH15+G4  &  -8.5& -4.12     & -8.73           & -8.68           & 0.05                 & -10.30            & 1.01              & -4.07           \\
			CNT+COOH20+G4  & 3.4 & -3.18     & -8.65           & -8.63           & 0.02                 & -9.38             & 1.23              & -4.62           \\
			CNT+COOH25+G5  & 5.8  & -5.55     & -8.83           & -8.62           & 0.21                 & -16.41            & 1.13              & -4.58           \\ \hline
		\end{tabular}
		\flushleft \tiny $^\dag$ Energies are in eV and $\Delta$RMSD is in \%.
	\end{center}

\end{table}

\begin{figure}[tbph]
	\centering
	\includegraphics[width=10cm]{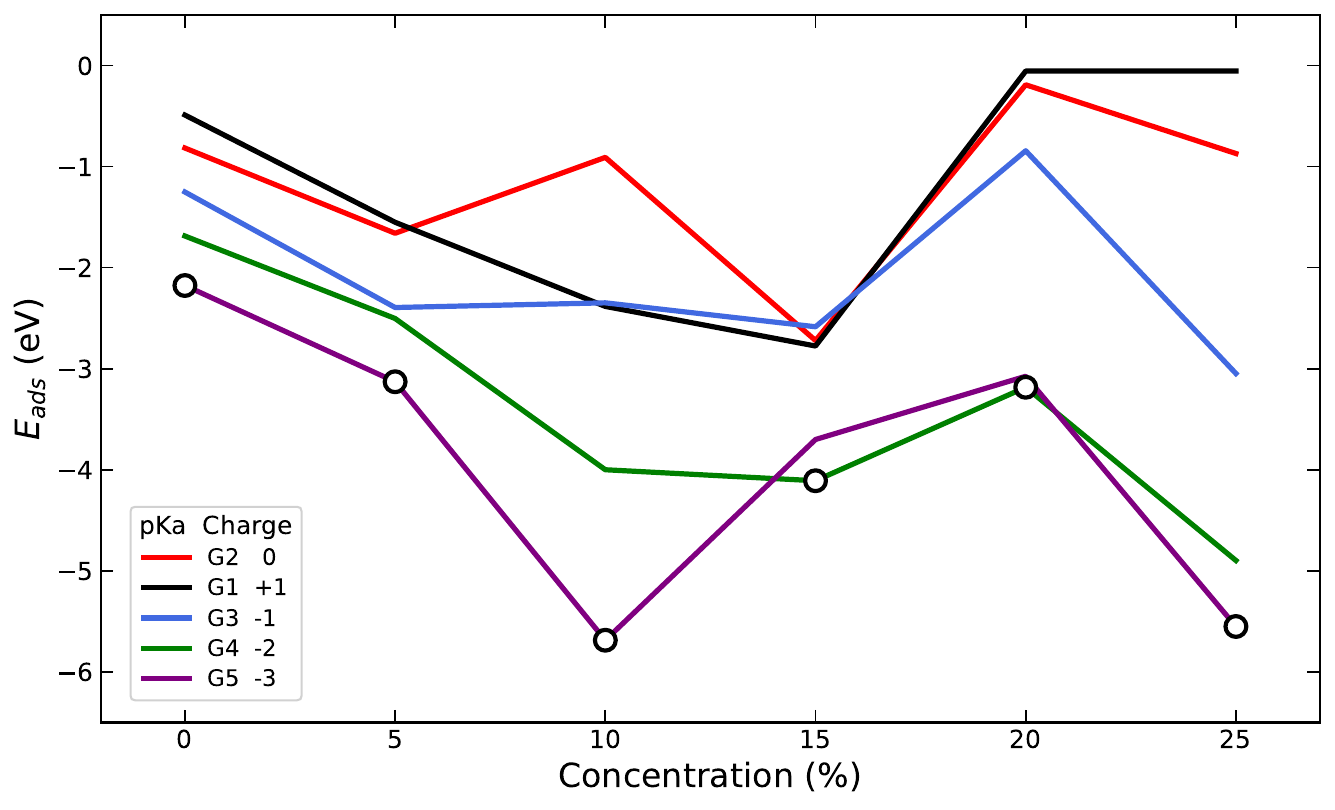}
	\caption{\label{Fig:Eads} Adsorption energy ($E_{ads}$, in eV) as a function of carboxyl functionalization degree (0-25\%) for the five glyphosate protonation states (G1-G5). Line colors denote the ionization form and its associated formal charge, as indicated in the legend (G1: +1; G2: 0; G3: -1; G4: -2; G5: -3). Open circles mark the representative complex selected at each concentration with more adsorption energy.)}
\end{figure}

The calculated adsorption energies for all glyphosate protonation states as a function of carboxyl content are shown in Figure~\ref{Fig:Eads}. In agreement with the trend reported for the hydroxyl-functionalized analog~\cite{Silva-ApplSurfSci-729-166060-2026}, the species (G4 and G5) consistently exhibit the most negative $E_{ads}$ across the whole concentration range, whereas the protonated and neutral forms (G1, G2) display the weakest binding, in several cases approaching zero at 20\% functionalization. This behavior reflects the higher negative charge density of the deprotonated forms, which strengthens electrostatic attraction and hydrogen bonding with the carboxyl adsorption sites~\cite{Dong-ChemEngJ-499-156654-2024}. 

The representative complex selected at each concentration (open circles in Figure~\ref{Fig:Eads}) follows a non-monotonic profile: adsorption strengthens from the pristine tube (CNT+G5, -2.18~eV) to a first maximum at CNT+COOH10+G5 (-5.69~eV), weakens at intermediate functionalization (CNT+COOH15+G4 and CNT+COOH20+G4, -4.11 and -3.18~eV), and strengthens again at CNT+COOH25+G5 (-5.55~eV). The local weakening near 20\%, most pronounced for the protonated forms (G4), indicates that beyond a certain carboxyl density the accumulation of neighboring COOH groups introduces steric and electrostatic competition that partially offsets the additional binding sites, rather than a simple linear intensification of the interaction. 
From a regeneration standpoint, this distribution is favorable: while the deprotonated representatives fall in the strong-chemisorption regime that favors capture but hinders desorption, the protonated forms at high carboxyl content (G1/G2 at 20\%-25\%) approach the moderate-interaction window associated with reversible adsorption and adsorbent reuse~\cite{Cui-Nanomaterials-13-2781-2023,Rahimi-SciRep-14-29282-2024}. The functionalized nanotube therefore spans both operating regimes depending on the ionization state of the pesticide, i.e. on the pH of the medium.
The frontier--orbital energies, gaps, and net charge transfer of the representative complexes are collected in Table~\ref{Tab:ElectRes}. 

The decomposition of the ALPB solvation free energy into its electrostatic ($\Delta G_{Elec}$), cavity/surface ($\Delta G_{SASA}$), and hydrogen--bond ($\Delta G_{HB}$) contributions (Table ~\ref{Tab:ElectRes}) clarifies the physical origin of the solvent stabilization. In all systems the electrostatic term dominates by roughly an order of magnitude, ranging from -9.38 eV (CNT+COOH20) to -16.41 eV (CNT+COOH25), which reflects the strong polarization coupling between the aqueous dielectric and the charged glyphosate--nanotube complex; its magnitude does not scale linearly with carboxyl content, being largest for the pristine tube (-14.35 eV) and the most functionalized G5 system (CNT+COOH25, -16.41 eV). The cavity term $\Delta G_{SASA}$ is small and, except for the pristine CNT (-0.21 eV), positive (destabilizing), increasing steadily with functionalization from 0.34 eV (COOH05) to 1.23 eV (CNT+COOH20); this growth is consistent with the larger and more corrugated solvent-accessible surface introduced by the COOH groups, which raises the free--energy cost of forming the solute cavity. The hydrogen--bond contribution $\Delta G_{HB}$ shows the clearest functionalization trend: it is essentially negligible for the pristine nanotube (-0.13 eV), where no carboxyl donors/acceptors are available, and becomes progressively more stabilizing as the carboxyl density increases, reaching -4.62 eV at CNT+COOH20. This monotonic strengthening of $\Delta G_{HB}$ directly quantifies the enhanced hydrogen bonding between the carboxyl groups and the surrounding water and mirrors the increase in explicit CNT--water and CNT--glyphosate hydrogen bonding expected upon functionalization. 

\begin{figure}[tbph]
	\centering
	\includegraphics[width=10cm]{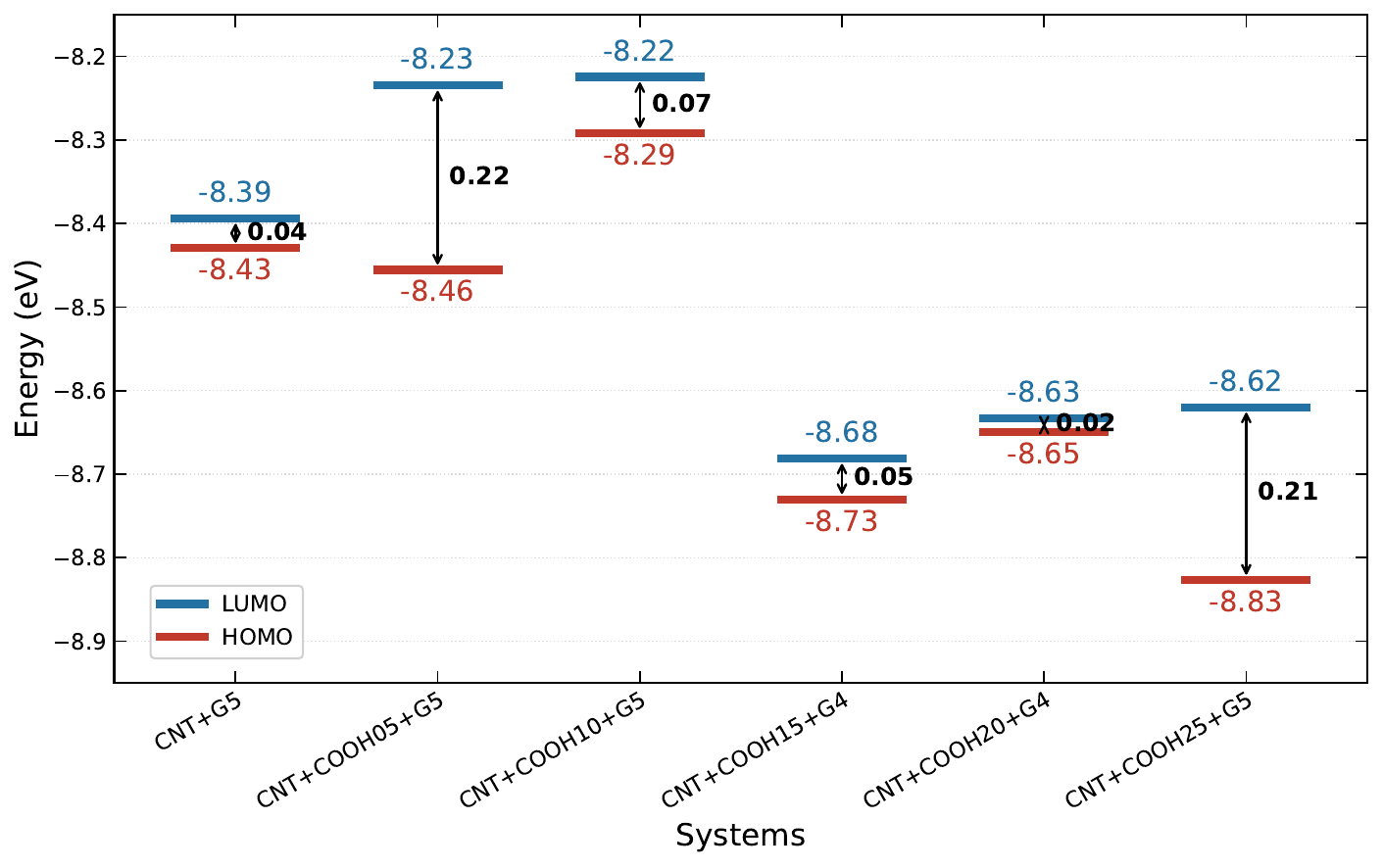}
	\caption{\label{Fig:Gaps} Frontier molecular orbital energy levels (in eV) for the representative CNT+COOH/glyphosate complexes in implicit water (ALPB). Red and blue bars indicate the HOMO
		and LUMO levels, respectively; the double-headed arrows and adjacent values report the
		HOMO-LUMO gap ($\Delta \varepsilon$) of each system.}
\end{figure}

All systems display very small HOMO--LUMO gaps (see Figure~\ref{Fig:Gaps}), ranging from 0.02 eV (CNT+COOH20) to 0.22 eV (CNT+COOH05), indicative of the near-degenerate frontier states characteristic of the metallic-like carbon-nanotube framework and consistent with a high electronic sensitivity toward adsorption. The gap does not vary monotonically with carboxyl content. More informative is the position of the frontier levels themselves: the weakly functionalized G5 systems (CNT, CNT+COOH05, CNT+COOH10) keep their HOMO in the -8.29 to -8.46 eV range, whereas the highly functionalized systems (CNT+COOH15, CNT+COOH20 and, most markedly, CNT+COOH25) are progressively stabilized, with the HOMO reaching -8.83 eV. This deepening of the occupied frontier level with increasing COOH density contrasts with the destabilization (less negative HOMO/LUMO) reported for the hydroxyl-functionalized systems, and is consistent with the electron-withdrawing character of the carboxyl group, which lowers the electron-donating capacity of the complex as functionalization increases~\cite{Milowska-JChemPhys-138-194704-2013,Lara-ChemPhys-428-117-2014}. 

\subsubsection*{Charge Transfer Analysis}
\begin{figure}[tbph]
	\centering
	\includegraphics[width=10cm]{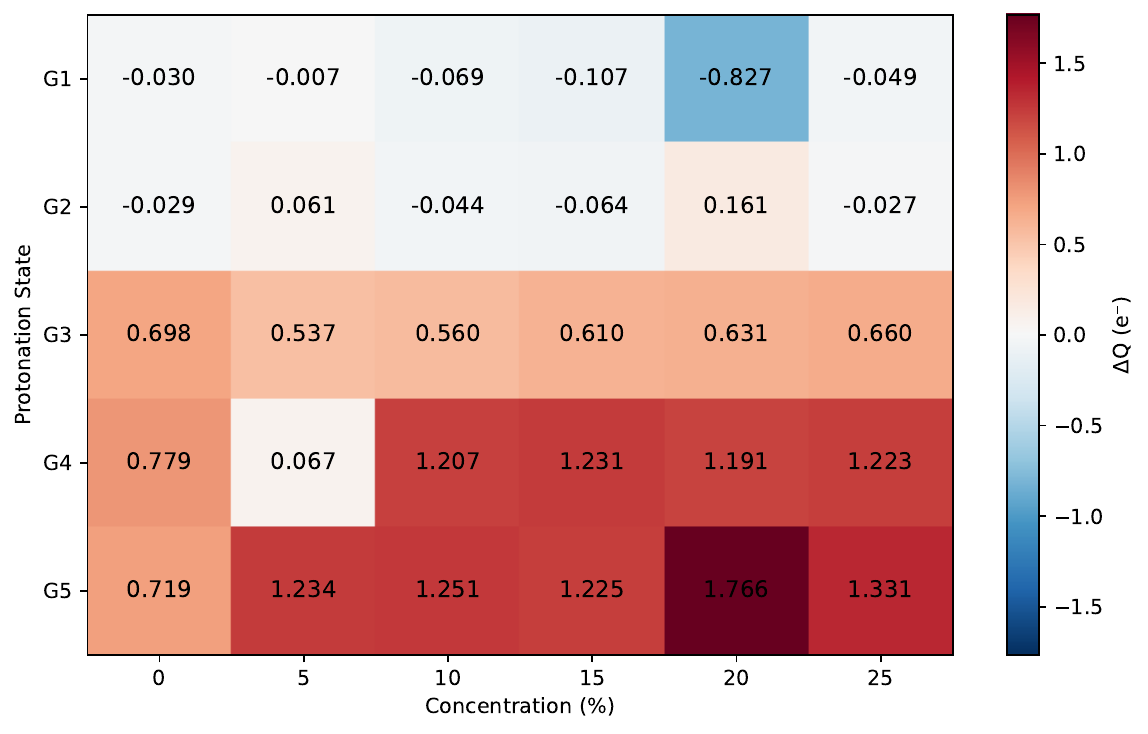}
	\caption{\label{Fig:Charges} Net charge transfer ($\Delta Q$, in electron charge ($e$)) between glyphosate and the functionalized hosts in implicit solvent (ON), represented as a heat map across the five protonation states (G1-G5) and six functionalization levels (CNT, COOH05-COOH25).}
\end{figure}

The net charge transfer $\Delta Q$ between glyphosate and the functionalized nanotubes in implicit solvent (Figure~\ref{Fig:Charges}), mapped across the five protonation states and six functionalization levels, shows that the protonation state of glyphosate is the primary factor governing the electronic interaction with the host. The species G4 and G5 exhibit the largest positive $\Delta Q$ reaching $\approx1.77e$ for CNT+COOH20+G5, which indicates that glyphosate donates a substantial amount of electron density to the nanotube.

This direction of transfer is consistent with the electron-rich nature of the deprotonated carboxylate groups and with the electron-withdrawing role of the carboxyl-functionalized surface inferred from the frontier-orbital analysis. The form G3 shows an intermediate and remarkably uniform donation ($\Delta Q \approx 0.54e$ across all hosts), while the G1 and G2 produce much smaller charge redistribution, with $\Delta Q$ values close to zero and a few localized cases of reversed transfer (host to glyphosate), most notably G1 on CNT+COOH20 ($\Delta Q \approx -0.83e$). The effect of carboxyl content is comparatively modest: increasing the COOH concentration produces only minor variations in $\Delta Q$ along each protonation row, in contrast to the pronounced dependence on ionization state. This indicates that, under aqueous conditions, the functional groups act primarily to organize adsorption and molecular recognition, whereas the electronic transduction,  the actual charge exchanged upon binding, is dictated mainly by the ionization state of the analyte.

\subsection{Topological analysis}
\label{Sec:Topo}

Topological analysis based on the Quantum Theory of Atoms in Molecules (QTAIM) was used to characterize the interactions responsible for stabilizing glyphosate on pristine carbon nanotubes (CNTs) and CNTs functionalized with carboxylic groups. Only \textbf{(3,-1)} bond critical points (BCPs), listed in Table S1, were considered because they correspond to a minimum in the electron density, $\rho(r)$, along the bond path and maxima in the two directions perpendicular to it. However, a bond path and BCP establish a topological connection between two attractors; they do not, by themselves, establish the presence of a covalent chemical bond~\cite{Bader-ChemRev-91-893-1991,Bader1994,Shahbazian-ChemEurJ-24-5401-2018,Popelier-JMolModel-28-276-2022}. This distinction is particularly important at adsorbent--adsorbate interfaces, where BCPs may be associated with dispersive and electrostatic contacts, hydrogen bonds, partially covalent interactions, or covalent bonding.

Accordingly, the chemical nature of each contact was assessed using a set of complementary descriptors: $\rho(r)$,  $\nabla^2\rho(r)$,  ELF, LOL, G(r), V(r), and H(r)~\cite{Becke-JChemPhys-92-5397-1990,Bader-ChemRev-91-893-1991,Bader1994,Schmider-JMolStructTHEOCHEM-527-51-2000,Lu-JComputChem-33-580-2012}. This combined approach is necessary because the sign of $\nabla^2\rho(r)$, considered alone, may be insufficient or misleading for contacts near the transition between closed-shell and shared-shell regimes~\cite{Rozas-JAmChemSoc-122-11154-2000,Grabowski-ChemRev-111-2597-2011}.

The sign of H(r) and the |V(r)|/G(r) ratio provide additional information on the local balance between potential and kinetic energy densities. In general, H(r)<0 indicates local predominance of potential energy and is compatible with an increased shared-electron contribution, whereas H(r)>0 is typical of interactions dominated by kinetic energy and commonly associated with closed-shell behavior~\cite{Rozas-JAmChemSoc-122-11154-2000,Grabowski-ChemRev-111-2597-2011}. In the present analysis, contacts with $\rho$>0.20~a.u. and $\nabla^2\rho$<0 were classified as predominantly covalent (shared--shell), whereas contacts with $\rho$<0.10~a.u. and $\nabla^2\rho$>0 were classified as closed-shell. For contacts near this boundary, the energetic descriptors were used as complementary criteria: H>0 and |V|/G<1 indicate closed-shell interactions; H<0 and 1<|V|/G<2 indicate interactions with partial covalent character; and H<0 and |V|/G>2 indicate predominantly shared-shell interactions~\cite{Rozas-JAmChemSoc-122-11154-2000,Grabowski-ChemRev-111-2597-2011}.

A total of 44 BCPs were identified across the six systems analyzed (Table S1). Of these, 41 exhibit H>0 and |V|/G<1 and were classified as closed-shell. Two BCPs show shared-shell character, with $\rho$>0.24~a.u., $\nabla^2\rho$<0, H<0, |V|/G > 2, and ELF>0.74. One BCP shows intermediate behavior, with H<0 and |V|/G=1.016, indicating a weak shared-electron contribution. The closed-shell contacts include O$\cdots$H/H$\cdots$O, O$\cdots$C, O$\cdots$O, H$\cdots$H, C$\cdots$H, O$\cdots$P, and O$\cdots$N interactions. Nevertheless, atomic contacts with the same topological classification should not be treated as energetically equivalent because they may involve different proportions of electrostatics, polarization, charge transfer, and dispersion~\cite{Grabowski-ChemRev-111-2597-2011,Shahbazian-ChemEurJ-24-5401-2018,Popelier-JMolModel-28-276-2022}.

\begin{figure}[tbph]
	\centering
	\subfigure[][CNT+G5]{\includegraphics[height=.2\textheight,keepaspectratio]{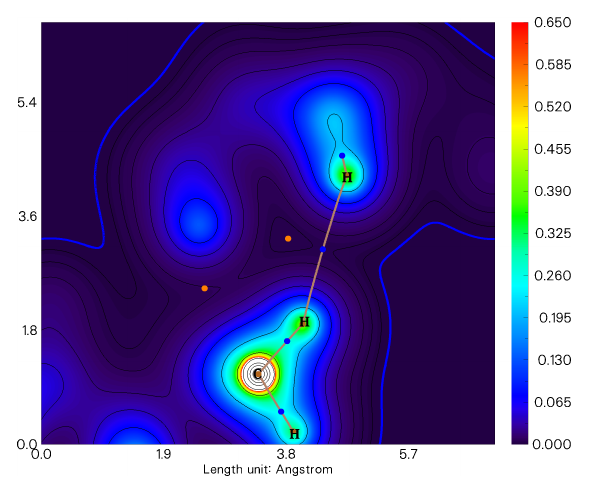}
		\label{Fig:RHO_CNT}}
	\subfigure[][CNT+COOH05+G5]{\includegraphics[height=.2\textheight,keepaspectratio]{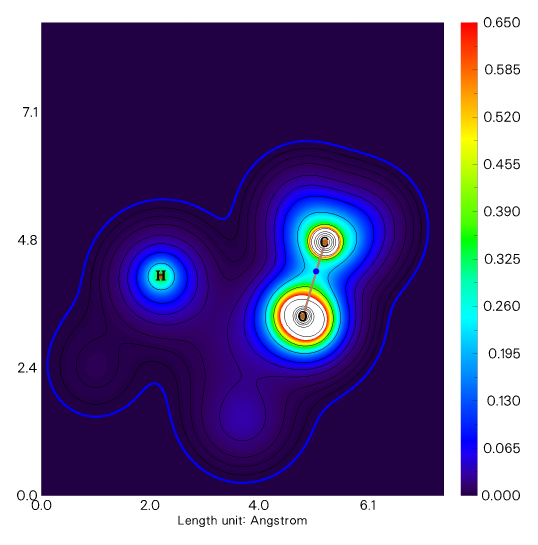}
		\label{Fig:RHO_OH05}}
	\subfigure[][CNT+COOH10+G5]{\includegraphics[height=.2\textheight,keepaspectratio]{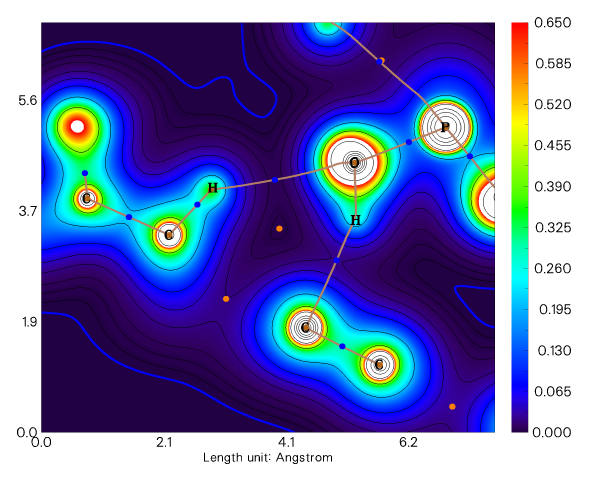}
		\label{Fig:RHO_OH10}}\\
	\subfigure[][CNT+COOH15+G4]{\includegraphics[height=.2\textheight,keepaspectratio]{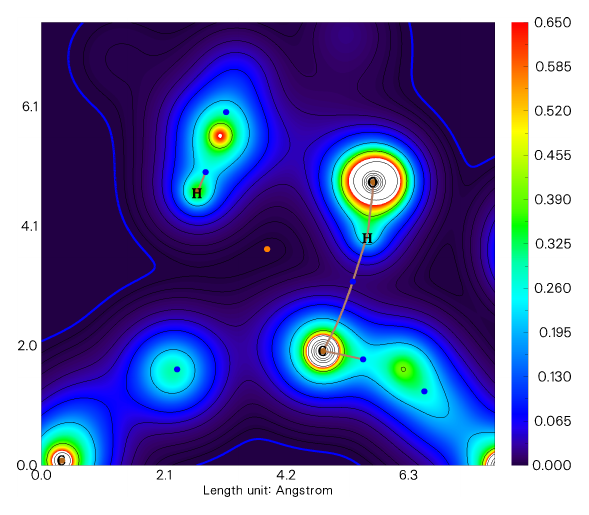}
		\label{Fig:RHO_OH15}}
	\subfigure[][CNT+COOH20+G4]{\includegraphics[height=.2\textheight,keepaspectratio]{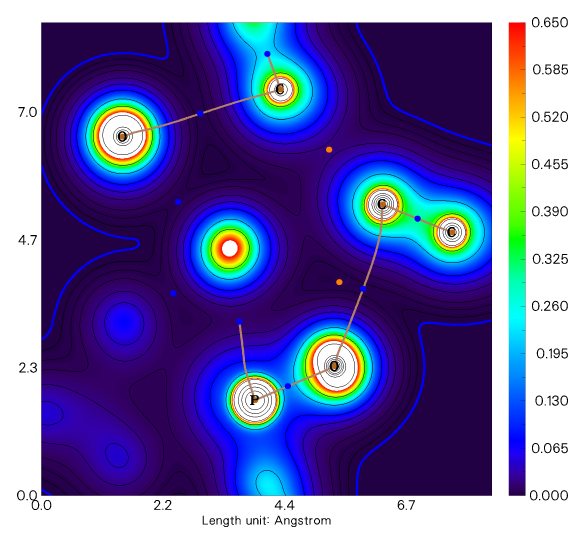}
		\label{Fig:RHO_OH20}}
	\subfigure[][CNT+COOH25+G5]{\includegraphics[height=.2\textheight,keepaspectratio]{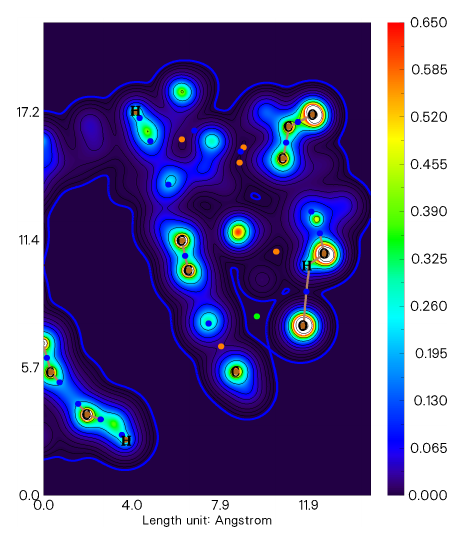}
		\label{Fig:RHO_OH25}}
	
		\caption{\label{Fig:EDens} Two-dimensional maps of the electron density, $\rho$.}
\end{figure}

The number of BCPs and their associated electron densities do not increase monotonically with the degree of carboxyl functionalization. Instead, functionalization reorganizes the adsorption interface. In CNT+G5 and CNT+COOH05+G5, stabilization is dominated by localized O--C contacts (BCPs 415 and 450, respectively), whereas the systems containing 10--25\% COOH develop multipoint interaction networks in which electron density is distributed among several weaker contacts. This reorganization is also evident in the electron-density maps (Figure~\ref{Fig:EDens}): CNT+G5 and CNT+COOH05+G5 show strongly concentrated density between O and C, while the more highly functionalized systems show several lower-intensity interaction domains distributed along the adsorbent--adsorbate interface.

In CNT+G5, the O142--C8 contact (BCP 415) has $\rho$=0.249~a.u., $\nabla^2\rho$=-0.355~a.u., H=-0.281~a.u., |V|/G=2.778, ELF=0.759, and LOL=0.639. These values identify a localized interaction with predominantly shared-shell character. The corresponding contact in CNT+COOH05+G5 (BCP 450, C139--O171) has comparable parameters ($\rho$=0.240~a.u., $\nabla^2\rho$=-0.288~a.u., H=-0.246~a.u., |V|/G=2.592, ELF=0.746, and LOL=0.631), indicating that low carboxyl-group coverage does not disrupt the localized anchoring mechanism. The high ELF and LOL values at both BCPs support substantial electron localization in these interatomic regions~\cite{Becke-JChemPhys-92-5397-1990,Schmider-JMolStructTHEOCHEM-527-51-2000}.

Beginning at 10\% functionalization, the topological organization of the interface changes. All BCPs in CNT+COOH10+G5, CNT+COOH15\-+G4, and CNT+COOH20\-+G4 lie in the closed-shell regime, with $\rho$ values no greater than 0.033~a.u. and with electron density distributed across 6--11 contacts per system (Table S1). This shift is consistent with a transition from localized anchoring to cooperative multipoint stabilization involving O$\cdots$C, O$\cdots$O, O$\cdots$H, H$\cdots$O, O$\cdots$P, and O$\cdots$N contacts. Such a framework indicates that several low- to moderate-density contacts can jointly orient and immobilize glyphosate on COOH-functionalized CNT surfaces. Their occurrence and relative magnitude depend on the local geometry and relative orientation of phosphonate, carboxylate, amino, and surface COOH groups, rather than solely on the nominal number of carboxylic groups~\cite{Espinosa-ChemPhysLett-285-170-1998,Grabowski-ChemRev-111-2597-2011}.

CNT+COOH15+G4 has the lowest summed BCP density among the functionalized systems (0.054~a.u., based on Table S1). This result indicates that the number of isolated carboxyl groups alone does not determine interfacial interaction strength. The simultaneous change from G5 to G4 and the resulting complex geometry also modulate the interfacial electron-density distribution. More generally, QTAIM descriptors at BCPs reflect the local electronic structure and geometry of each interaction; they should not be interpreted as direct functions of global chemical composition or nominal surface-group coverage alone~\cite{Bader-ChemRev-91-893-1991,Bader1994,Espinosa-ChemPhysLett-285-170-1998}.

\begin{figure}[tbph]
	\centering
	\subfigure[][CNT+G5]{\includegraphics[height=.2\textheight,keepaspectratio]{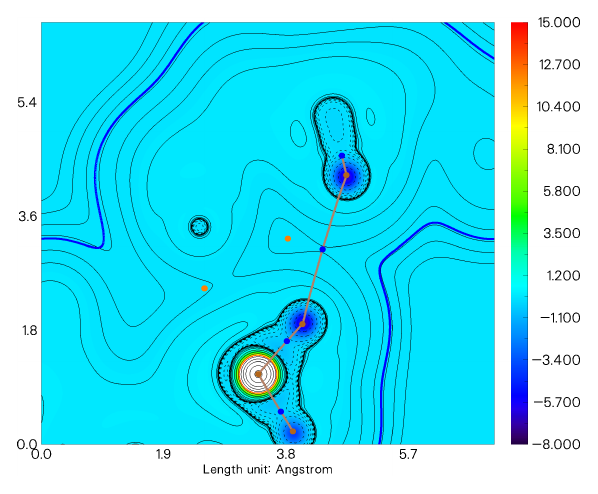}
		\label{Fig:LAP_CNT}}
	\subfigure[][CNT+COOH05+G5]{\includegraphics[height=.2\textheight,keepaspectratio]{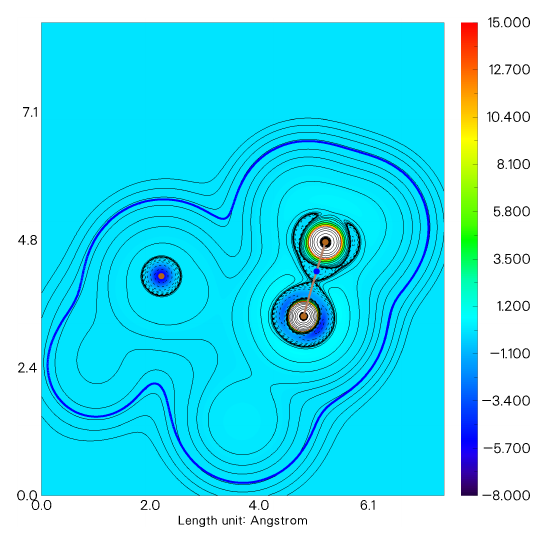}
		\label{Fig:LAP_OH05}}
	\subfigure[][CNT+COOH10+G5]{\includegraphics[height=.2\textheight,keepaspectratio]{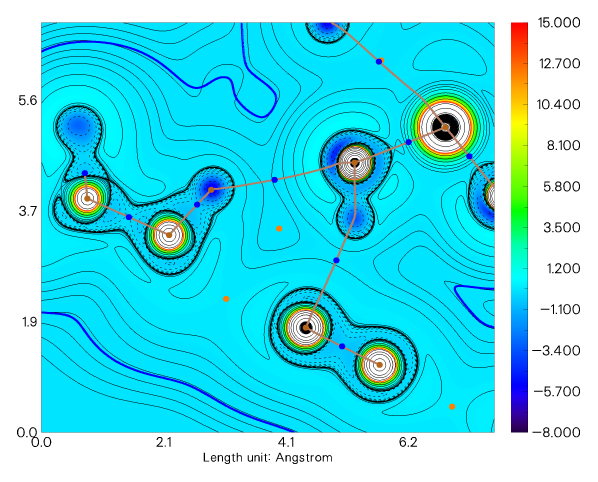}
		\label{Fig:LAP_OH10}}\\
	\subfigure[][CNT+COOH15+G4]{\includegraphics[height=.2\textheight,keepaspectratio]{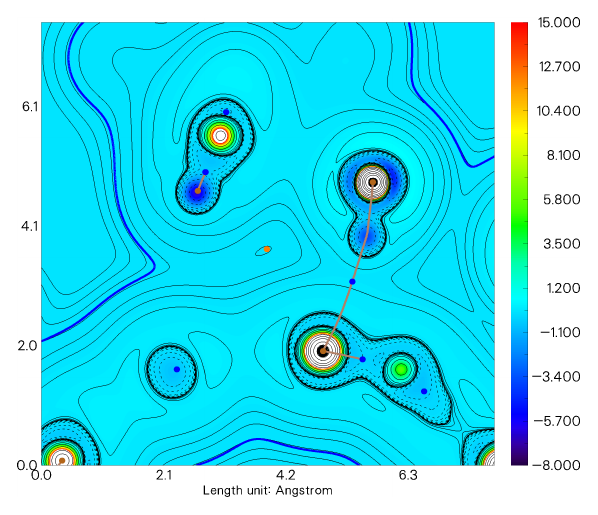}
		\label{Fig:LAP_OH15}}
	\subfigure[][CNT+COOH20+G4]{\includegraphics[height=.2\textheight,keepaspectratio]{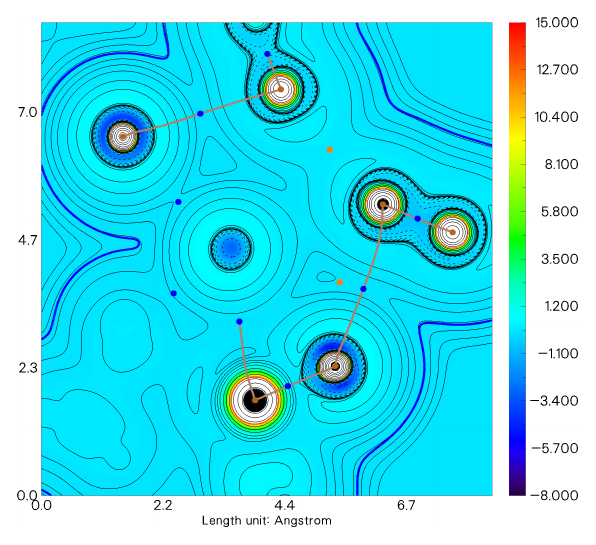}
		\label{Fig:LAP_OH20}}
	\subfigure[][CNT+COOH25+G5]{\includegraphics[height=.2\textheight,keepaspectratio]{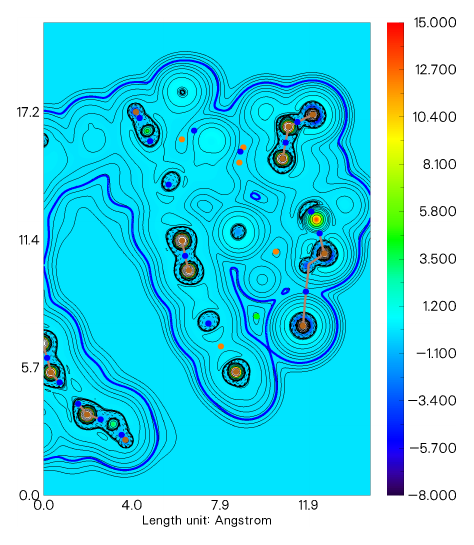}
		\label{Fig:LAP_OH25}}
	
	\caption{\label{Fig:Lapla} Two-dimensional maps of the Laplacian of the electron density, $\nabla^2\rho$.}
\end{figure}

In CNT+COOH20+G4, two O$\cdots$O contacts (BCPs 205 and 647; $\rho$=0.033 and 0.031~a.u.; |V|/G=0.928 and 0.915, respectively) define a double-anchoring motif, supplemented by nine secondary contacts with $\rho$ values between 0.001 and 0.012~a.u. All of these contacts are consistent with closed-shell interactions. By contrast, CNT+COOH25+G5 has a distinct electronic organization, with a summed BCP density approximately 36\% higher than that of CNT+COOH20+G4 and the only BCP with partial covalent character (BCP 311, O284$\cdots$O197; $\rho$=0.058~a.u., $\nabla^2\rho$=0.214~a.u., H=-0.001~a.u., |V|/G=1.016, ELF=0.178, and LOL=0.318). Although $\nabla^2\rho$ remains positive (Figure~\ref{Fig:Lapla}), the negative H value and |V|/G slightly above unity place this contact in the intermediate region between closed-shell and shared-shell regimes. This interpretation is consistent with QTAIM analyses in which stronger interactions may retain a positive Laplacian while showing H<0, indicating a non-negligible shared-electron contribution~\cite{Rozas-JAmChemSoc-122-11154-2000,Grabowski-ChemRev-111-2597-2011}.

\begin{figure}[tbph]
	\centering
	\subfigure[][CNT+G5]{\includegraphics[height=.2\textheight,keepaspectratio]{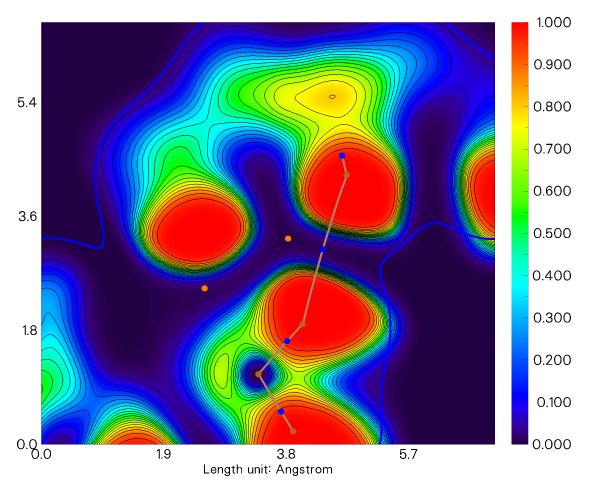}
		\label{Fig:ELFL_CNT}}
	\subfigure[][CNT+COOH05+G5]{\includegraphics[height=.2\textheight,keepaspectratio]{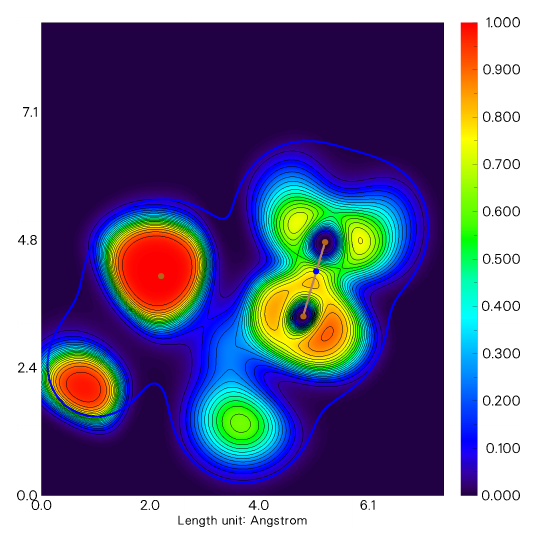}
		\label{Fig:ELF_OH05}}
	\subfigure[][CNT+COOH10+G5]{\includegraphics[height=.2\textheight,keepaspectratio]{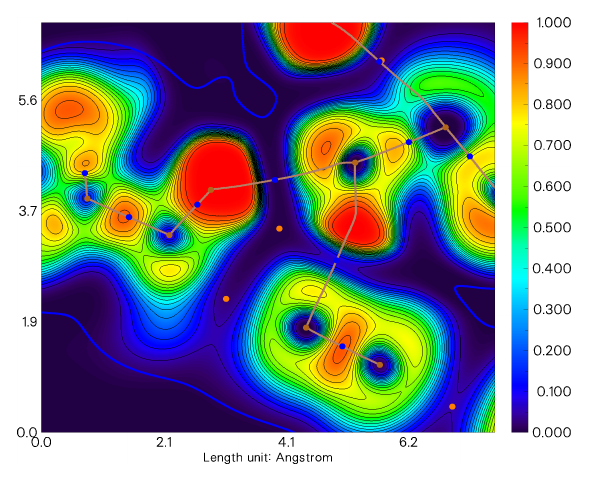}
		\label{Fig:ELF_OH10}}\\
	\subfigure[][CNT+COOH15+G4]{\includegraphics[height=.2\textheight,keepaspectratio]{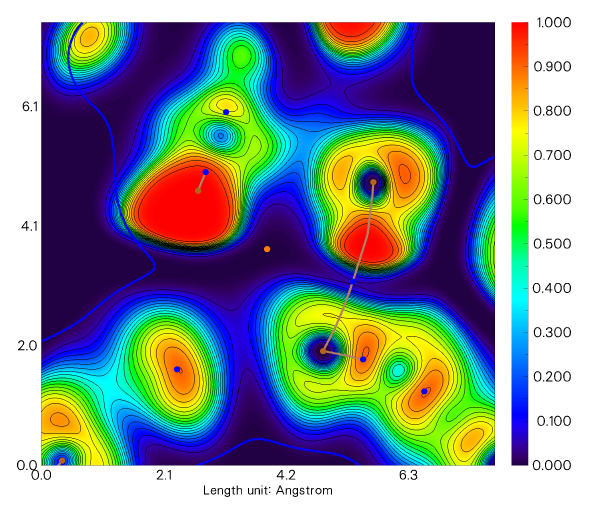}
		\label{Fig:ELF_OH15}}
	\subfigure[][CNT+COOH20+G4]{\includegraphics[height=.2\textheight,keepaspectratio]{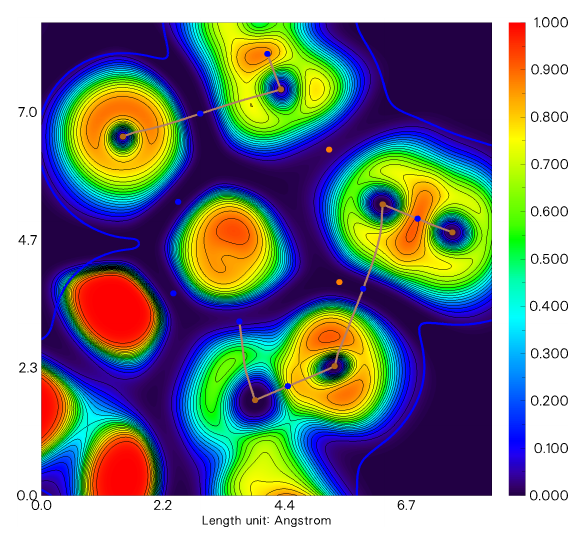}
		\label{Fig:ELF_OH20}}
	\subfigure[][CNT+COOH25+G5]{\includegraphics[height=.2\textheight,keepaspectratio]{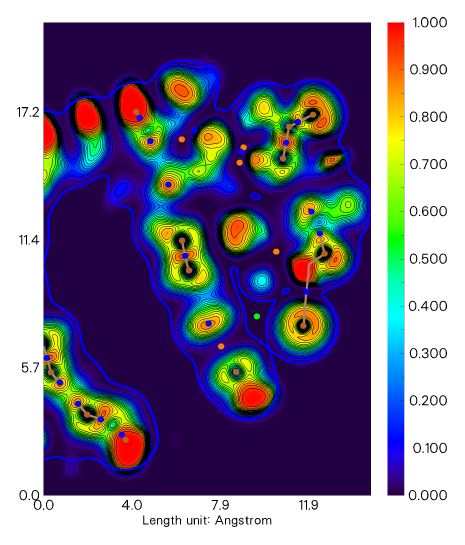}
		\label{Fig:ELF_OH25}}
	
	\caption{\label{Fig:ELF} Two-dimensional maps of the electronic locatization function, ELF.}
\end{figure}

\begin{figure}[tbph]
	\centering
	\subfigure[][CNT+G5]{\includegraphics[height=.2\textheight,keepaspectratio]{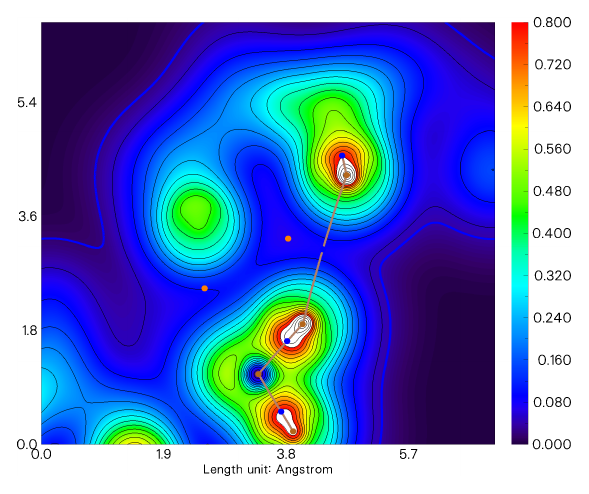}
		\label{Fig:LOL_CNT}}
	\subfigure[][CNT+COOH05+G5]{\includegraphics[height=.2\textheight,keepaspectratio]{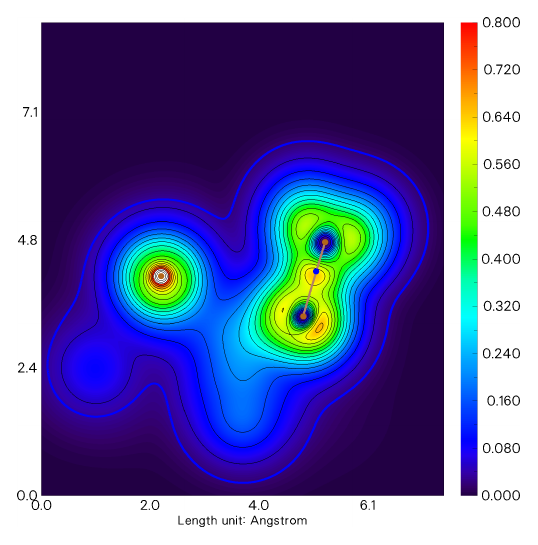}
		\label{Fig:LOL_OH05}}
	\subfigure[][CNT+COOH10+G5]{\includegraphics[height=.2\textheight,keepaspectratio]{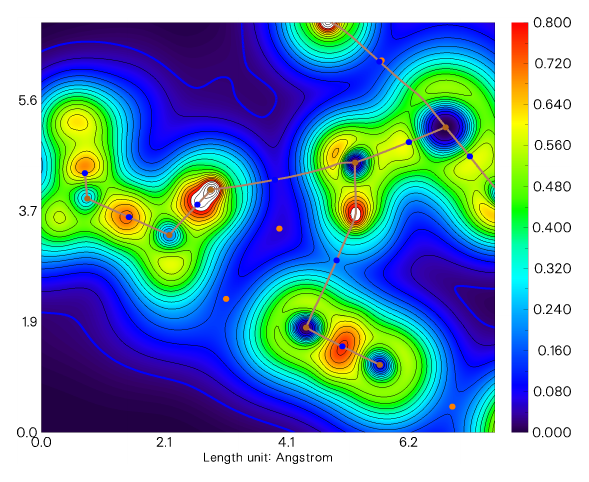}
		\label{Fig:LOL_OH10}}\\
	\subfigure[][CNT+COOH15+G4]{\includegraphics[height=.2\textheight,keepaspectratio]{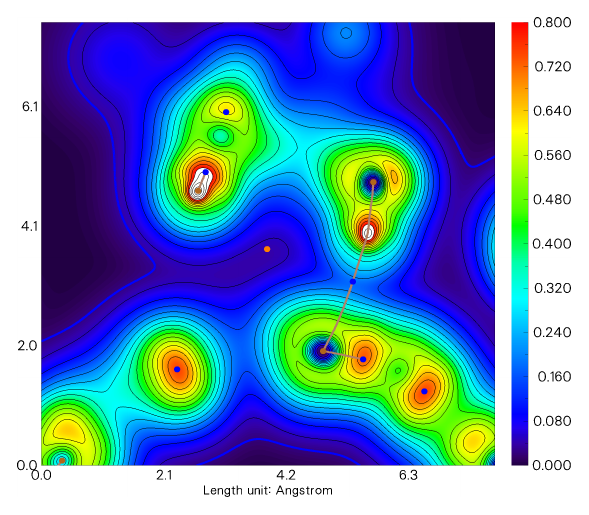}
		\label{Fig:LOL_OH15}}
	\subfigure[][CNT+COOH20+G4]{\includegraphics[height=.2\textheight,keepaspectratio]{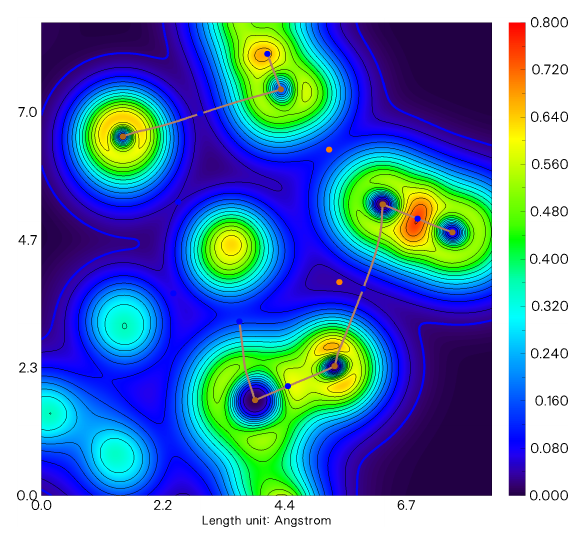}
		\label{Fig:LOL_OH20}}
	\subfigure[][CNT+COOH25+G5]{\includegraphics[height=.2\textheight,keepaspectratio]{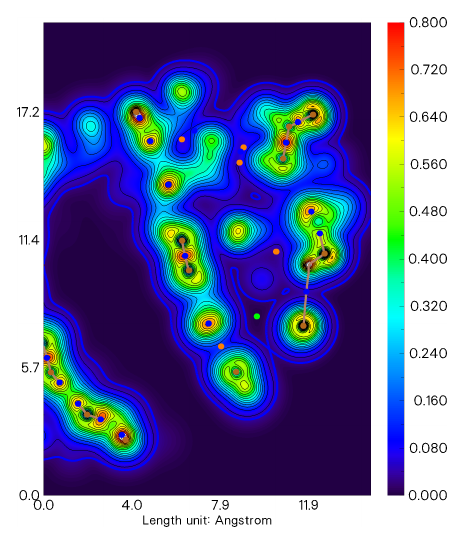}
		\label{Fig:LOL_OH25}}
	
	\caption{\label{Fig:LOL} Two-dimensional maps of the localized orbital locator, LOL.}
\end{figure}

The ELF and LOL values corroborate the classification based on QTAIM energy descriptors (Figures~\ref{Fig:ELF} and~\ref{Fig:LOL}) by independently depicting electron localization in the interatomic regions. BCPs 415 in CNT+G5 and 450 in CNT+COOH05+G5, both classified as shared-shell, display ELF values above 0.74 and LOL values above 0.63, consistent with substantial electron localization and the negative $\nabla^2\rho$ and H values observed at these BCPs. In contrast, closed-shell BCPs display ELF values of up to 0.077 and LOL values of up to 0.224, indicating low electron localization in the interatomic region. BCP 311 in CNT+COOH25+G5 displays intermediate values (ELF=0.178 and LOL=0.318), consistent with its polarized and weakly shared character~\cite{Becke-JChemPhys-92-5397-1990,Lu-JComputChem-33-580-2012,Schmider-JMolStructTHEOCHEM-527-51-2000} .

\subsection{Molecular Dynamics Simulations}
\label{MolDyn}

Molecular dynamics (MD) simulations were carried out on the energetically optimized geometries of all complexes (Figure~\ref{Fig:STRUCALL}A). The results revealed distinct binding behaviors, corroborating what had already been observed in the analysis of the electronic and topological properties. To quantify the structural stability of the complexes, the radial distribution functions (RDFs), the minimum distance between glyphosate and the nanotube backbone, and the number of hydrogen bonds (HBonds) were computed over the course of the simulation using the VMD software~\cite{vmd}.

\begin{figure}[tbph]
	\centering
	\includegraphics[width=\textwidth]{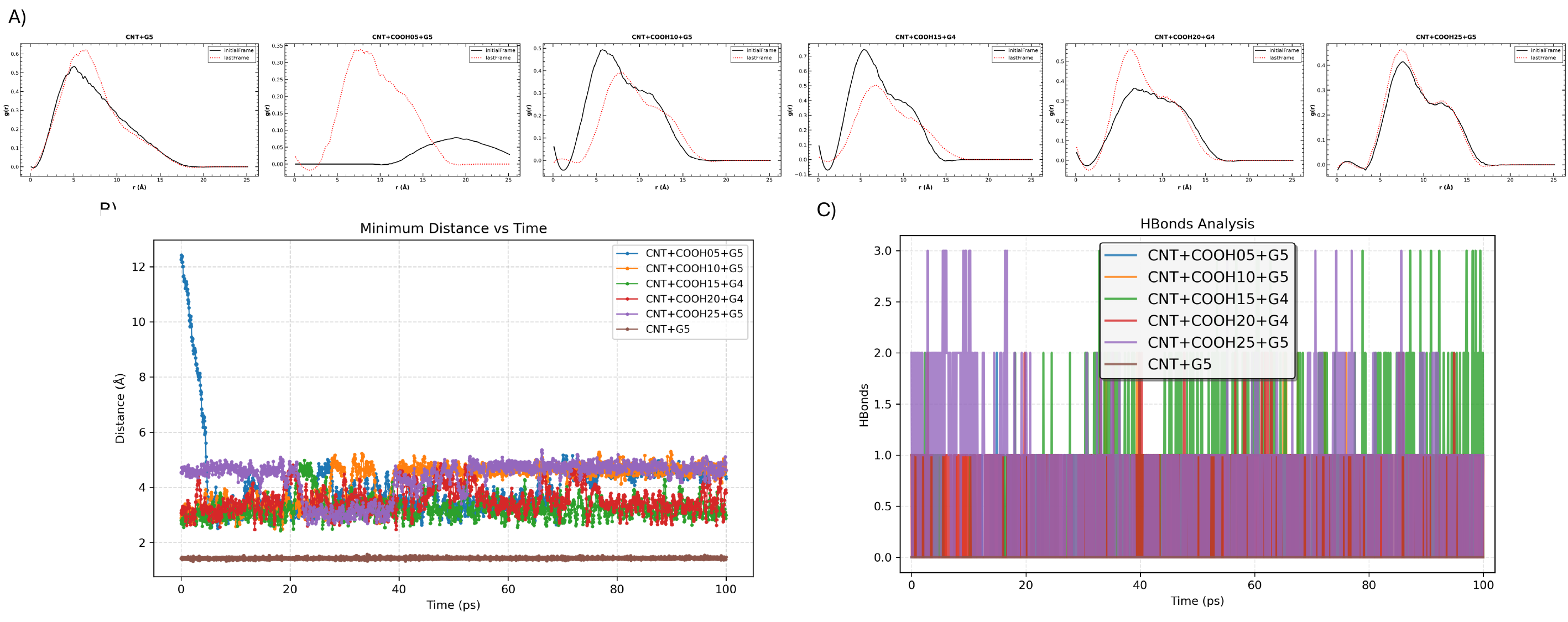}
	\caption{\label{Fig:MolDyn} Panel A: Radial distribution function for initial (solid black line) and last (dashed red line) frames. Panel B: Minimal distance between the nanotubes and glyphosate during the simulation time. Panel C: Hydrogen bonds. Calculations done using VMD software~\cite{vmd}.}
\end{figure}

The collective spatial distribution of the molecular interactions was examined through RDFs computed for systems in which glyphosate, at each degree of ionization, interacts with the nanotube, thereby giving access to the relative probability of finding the glyphosate molecule at a distance $r$ from the CNT surface. Figure~\ref{Fig:MolDyn}A presents these calculations, with the solid black line corresponding to the initial frame and the dashed red line to the final frame. With the exception of the CNT+COOH05+G5 system, the initial and final configurations of GLY are found at comparable distances, between 5--8~\AA~from the nanotubes. A closer inspection of the initial configuration of CNT+COOH05+G5 (see Figure~\ref{Fig:STRUCALL}A(b)) reveals that this is precisely the arrangement in which GLY sits farthest from the nanotube ($\sim$19~\AA); once the simulation is performed, however, its final configuration coincides with that of all other systems. Since in no case does GLY drift away from the nanotube by the end of the trajectory, a favorable interaction between the two can be confirmed.

This picture is reinforced by Figure~\ref{Fig:MolDyn}B, which shows the minimum distance between glyphosate and the nanotubes. During the first stages of the simulation ($t < 10$~ps), GLY approaches the nanotube surface; thereafter, its separation from CNT+COOH05 settles within 2--5~\AA, an interval that likewise holds for all remaining systems.

As discussed in Section~\ref{ElecStruc}, hydrogen--bond interactions play an important role in stabilizing the complexes. Figure~\ref{Fig:MolDyn}C shows the number of hydrogen bonds (HBonds) formed between the functionalized carbon nanotubes (CNTs) and glyphosate as a function of simulation time. As expected, the pristine CNT system formed no HBonds at all, a consequence of its surface lacking donor and acceptor groups. All functionalized CNT systems, by contrast, sustained at least two HBonds throughout the trajectory. Two of them, CNT+COOH15+G4 (green line) and CNT+COOH25+G5 (purple line), stood out for the larger number of HBonds formed, whereas the remaining complexes, although limited to two, preserved them consistently over the entire simulation.

\section{CONCLUSIONS}
\label{Sec:Conc}
The present study demonstrates that carboxyl--functionalized (10,0) carbon nanotubes provide a tunable platform for glyphosate capture, with adsorption strength primarily governed by the herbicide’s ionization state. Deprotonated forms G4 and G5 consistently exhibit the most negative adsorption energies, while G1 and G2 display weak binding that approaches reversible adsorption at high COOH loadings.

The non-monotonic dependence of $E_{ads}$ on carboxyl content, with maxima for G5 at 10\% and 25\% and a local minimum near 20\%, reveals a competition between new adsorption sites and steric/electrostatic crowding of neighboring groups. Charge transfer reveal that the dominant design variable is glyphosate protonation state: deprotonated forms donate up to $\sim$1.8$e$ to the nanotube and bind strongly, while protonated/neutral forms contribute little charge and exhibit weaker adsorption. Solvation analysis shows that electrostatic stabilization dominates, whereas hydrogen-bond contributions grow steadily with functionalization, reflecting increasingly cooperative CNT--water and CNT--glyphosate networks.

Molecular dynamics confirms that all complexes remain bound over 100~ps, with functionalized CNTs sustaining multiple hydrogen bonds while pristine tubes rely solely on non-hydrogen-bonding contacts. These results underscore that pH and functionalization jointly control both structural and electronic stabilization, offering molecular-level guidelines for designing CNT--based glyphosate sensors and adsorbents.

\section*{Acknowledgements}
I.C. is grateful to the Brazilian funding agency CNPq for the research scholarship (304937\allowbreak/2023-1). This study was financed in part by the Coordena\c{c}\~ao de Aperfei\c{c}oamento de Pessoal de N\'{\i}vel Superior-Brasil (CAPES)-Finance code 001. Part of the results presented here were developed with the help of CENAPAD-SP (Centro Nacional de Processamento de Alto Desempenho em S\~ao Paulo) grant UNICAMP/FINEP-MCT, and the National Laboratory for Scientific Computing (LNCC/MCTI, Brazil) for providing HPC resources of the Santos Dumont supercomputer and CENAPAD-UFC (Centro Nacional de Processamento de Alto Desempenho, at Universidade Federal do Cear\'a).

\section*{CRediT authorship contribution statement}
\textbf{H.~T.~Silva}: Formal analysis, Investigation, Writing-original draft, Writing-review \& editing. \textbf{L.~C.~S.~Faria}: Formal analysis, Investigation, Writing-original draft, Writing-review \& editing. \textbf{C.~Aguiar}: Formal analysis, Investigation, Writing-original draft, Writing-review \& editing.
\textbf{T.~A.~Aversi-Ferreira}: Formal analysis, Investigation, Writing-original draft, Writing-review \& editing. \textbf{I.~Camps}: Conceptualization, Formal analysis, Methodology, Project administration, Resources, Software, Supervision, Writing-review \& editing.

\section*{Declaration of competing interest}

The authors declare that they have no known competing financial interests or personal relationships that could have appeared to influence the work reported in this paper.

\section*{Data availability}
\label{data_avail}
The raw data required to reproduce these findings are available to download from Zenodo repository.
\newpage
\bibliographystyle{elsarticle-num}
\bibliography{biblio}

@Article{iijima-Nature-363-603,
  author    = {S. Iijima and T. Ichihashi},
  doi       = {10.1038/363603a0},
  journal   = {Nature},
  pages     = {603--605},
  title     = {Single-shell carbon nanotubes of 1-nm diameter},
  volume    = {363},
  year      = {1993},
}

@Article{Gill_2017,
  author           = {Gill, Jatinder Pal Kaur and Sethi, Nidhi and Mohan, Anand and Datta, Shivika and Girdhar, Madhuri},
  doi              = {10.1007/s10311-017-0689-0},
  journal          = {Environ. Chem. Lett.},
  pages            = {401--426},
  title            = {Glyphosate toxicity for animals},
  volume           = {16},
  year             = {2017},
  creationdate     = {2025-12-15T09:49:32},
  issn             = {1610-3661},
  modificationdate = {2025-12-15T09:50:12},
  publisher        = {Springer Science and Business Media LLC},
}

@Article{Singh_2024,
  author           = {Singh, Reenu and Shukla, Akanksha and Kaur, Gurdeep and Girdhar, Madhuri and Malik, Tabarak and Mohan, Anand},
  doi              = {10.1021/acsomega.3c08080},
  journal          = {ACS Omega},
  pages            = {6165--6183},
  title            = {Systemic analysis of glyphosate impact on environment and human health},
  volume           = {9},
  year             = {2024},
  creationdate     = {2025-12-15T09:52:57},
  issn             = {2470-1343},
  modificationdate = {2025-12-15T09:53:50},
  publisher        = {American Chemical Society (ACS)},
}

@Article{Evalen_2024,
  author           = {Evalen, P. S. and Barnhardt, E. N. and Ryu, J. and Stahlschmidt, Z. R.},
  doi              = {10.1016/j.envpol.2024.123669},
  journal          = {Environ. Pollut.},
  pages            = {123669},
  title            = {Toxicity of glyphosate to animals: A meta-analytical approach},
  volume           = {347},
  year             = {2024},
  creationdate     = {2025-12-15T09:51:14},
  issn             = {0269-7491},
  modificationdate = {2025-12-15T09:51:54},
  publisher        = {Elsevier BV},
}

@Article{Bannwarth_2020,
  author           = {Bannwarth, Christoph and Caldeweyher, Eike and Ehlert, Sebastian and Hansen, Andreas and Pracht, Philipp and Seibert, Jakob and Spicher, Sebastian and Grimme, Stefan},
  doi              = {10.1002/wcms.1493},
  journal          = {WIREs Comput. Mol. Sci.},
  pages            = {e1493},
  title            = {Extended tight-binding quantum chemistry methods},
  volume           = {11},
  year             = {2020},
  creationdate     = {2025-08-19T11:03:17},
  issn             = {1759-0884},
  modificationdate = {2025-11-03T09:14:41},
  publisher        = {Wiley},
}

@Article{xTB_GFN2,
  author           = {C. Bannwarth and S. Ehlert and S. Grimme},
  doi              = {10.1021/acs.jctc.8b01176},
  journal          = {J. Chem. Theory Comput.},
  pages            = {1652--1671},
  title            = {{GFN2--xTB}--{An} accurate and broadly parametrized
		  self-consistent tight-binding quantum chemical method with
		  multipole electrostatics and density-dependent dispersion
		  contributions},
  volume           = {15},
  year             = {2019},
  creationdate     = {2022-12-27T20:22:38},
  modificationdate = {2023-07-03T17:22:10},
  publisher        = {American Chemical Society ({ACS})},
}

@Article{Schlegel_2011,
  author           = {Schlegel, H. Bernhard},
  doi              = {10.1002/wcms.34},
  journal          = {WIREs Comput. Mol. Sci.},
  pages            = {790--809},
  title            = {Geometry optimization},
  volume           = {1},
  year             = {2011},
  creationdate     = {2025-08-27T10:52:36},
  issn             = {1759-0884},
  modificationdate = {2025-08-27T14:52:24},
  publisher        = {Wiley},
}

@Article{Aguiar_2024,
  author           = {Aguiar, C. and Camps, I.},
  doi              = {10.1016/j.surfin.2024.104874},
  journal          = {Surfaces and Interfaces},
  pages            = {104874},
  title            = {Exploring the potential of boron-nitride nanobelts in environmental applications: Greenhouse gases capture},
  volume           = {52},
  year             = {2024},
  creationdate     = {2025-08-19T11:10:51},
  issn             = {2468-0230},
  modificationdate = {2025-11-28T15:01:32},
  publisher        = {Elsevier BV},
}

@Article{Herath_2019,
  author           = {G. A. D. Herath and L. S. Poh and W. J. Ng},
  doi              = {10.1016/j.chemosphere.2019.04.078},
  journal          = {Chemosphere},
  pages            = {533--540},
  title            = {Statistical optimization of glyphosate adsorption by biochar and activated carbon with response surface methodology},
  volume           = {227},
  year             = {2019},
  creationdate     = {2025-08-19T12:28:23},
  issn             = {0045-6535},
  modificationdate = {2025-11-28T15:01:32},
  publisher        = {Elsevier BV},
}

@Article{xTB-dock,
  author           = {C. Plett and S. Grimme},
  doi              = {10.1002/anie.202214477},
  journal          = {Angew. Chem. Int. Ed.},
  title            = {Automated and efficient generation of general molecular aggregate structures},
  volume           = {62},
  year             = {2022},
  creationdate     = {2023-01-18T08:26:59},
  modificationdate = {2025-11-28T15:01:32},
  publisher        = {Wiley},
}

@Article{vmd,
  author           = {William Humphrey and Andrew Dalke and Klaus Schulten},
  doi              = {10.1016/0263-7855(96)00018-5},
  journal          = {J. Mol. Graph.},
  pages            = {33--38},
  title            = {{VMD}: Visual molecular dynamics},
  volume           = {14},
  year             = {1996},
  creationdate     = {2022-12-31T13:33:17},
  modificationdate = {2024-04-19T21:33:09},
  publisher        = {Elsevier {BV}},
}

@Article{Mewes_2021,
  author           = {Mewes, Jan-Michael and Hansen, Andreas and Grimme, Stefan},
  doi              = {10.1002/anie.202102679},
  journal          = {Angew. Chem. - Int. Ed.},
  pages            = {13144--13149},
  title            = {{Comment on ``The Nature of Chalcogen-Bonding-Type Tellurium--Nitrogen Interactions'': Fixing the Description of Finite-Temperature Effects Restores the Agreement Between Experiment and Theory}},
  volume           = {60},
  year             = {2021},
  creationdate     = {2025-12-01T17:33:01},
  issn             = {1521-3773},
  modificationdate = {2025-12-01T17:39:45},
  publisher        = {Wiley},
}

@Article{charges_CM5,
  author           = {Marenich, A. V. and Jerome, S. V. and Cramer, C. J. and Truhlar, D. G.},
  doi              = {10.1021/ct200866d},
  journal          = {J. Chem. Theory Comput.},
  pages            = {527--541},
  title            = {{Charge Model 5: an extension of Hirshfeld population analysis for the accurate description of molecular interactions in gaseous and condensed phases}},
  volume           = {8},
  year             = {2012},
  creationdate     = {2023-02-16T11:15:08},
  modificationdate = {2025-12-10T15:45:43},
  publisher        = {American Chemical Society ({ACS})},
}

@Article{Kohn_2023,
  author           = {Kohn, J. T. and Gildemeister, N. and Grimme, S. and Fazzi, D. and Hansen, A.},
  doi              = {10.1063/5.0167484},
  journal          = {J. Chem. Phys.},
  pages            = {144106},
  title            = {Efficient calculation of electronic coupling integrals with the dimer projection method via a density matrix tight-binding potential},
  volume           = {159},
  year             = {2023},
  creationdate     = {2025-08-20T14:44:33},
  issn             = {1089-7690},
  modificationdate = {2025-11-28T15:01:32},
  publisher        = {AIP Publishing},
}

@Article{bsse,
  author    = {S. F. Boys and F. Bernardi},
  doi       = {10.1080/00268977000101561},
  journal   = {Mol. Phys.},
  pages     = {553--566},
  title     = {{The calculation of small molecular interactions by the
		  differences of separate total energies. Some procedures
		  with reduced errors}},
  volume    = {19},
  year      = {1970},
}

@Article{Multiwfn2,
  author           = {Lu, Tian},
  doi              = {10.1063/5.0216272},
  journal          = {J. Chem. Phys.},
  pages            = {082503},
  title            = {A comprehensive electron wavefunction analysis toolbox for chemists, {Multiwfn}},
  volume           = {161},
  year             = {2024},
  creationdate     = {2025-08-20T15:00:34},
  issn             = {1089-7690},
  modificationdate = {2025-11-28T15:01:32},
  publisher        = {AIP Publishing},
}

@Book{Bader1994,
  author           = {Bader, Richard F. W.},
  publisher        = {Clarendon Press},
  title            = {Atoms in molecules: a quantum theory},
  year             = {1994},
  address          = {Oxford},
  series           = {International series of monographs on chemistry},
  modificationdate = {2026-08-28T10:24:50},
}

@Article{Koch_2024,
  author           = {Koch, Daniel and Pavanello, Michele and Shao, Xuecheng and Ihara, Manabu and Ayers, Paul W. and Matta, Ch{\'{e}}rif F. and Jenkins, Samantha and Manzhos, Sergei},
  doi              = {10.1021/acs.chemrev.4c00297},
  journal          = {Chem. Rev.},
  pages            = {12661--12737},
  title            = {The analysis of electron densities: from basics to emergent applications},
  volume           = {124},
  year             = {2024},
  creationdate     = {2025-08-20T15:41:23},
  issn             = {1520-6890},
  modificationdate = {2025-11-03T09:14:41},
  publisher        = {American Chemical Society (ACS)},
}

@Article{Fedorov_2025,
  author           = {Fedorov, Igor},
  doi              = {10.3390/ma18081790},
  journal          = {Materials},
  pages            = {1790},
  title            = {Topological analysis of electron density in graphene/benzene and graphene/{hBN}},
  volume           = {18},
  year             = {2025},
  creationdate     = {2025-08-20T15:43:40},
  issn             = {1996-1944},
  modificationdate = {2025-11-28T15:01:32},
  publisher        = {MDPI AG},
}

@Article{Martinez_2003,
  author           = {Mart\'{\i}nez, J. M. and Mart\'{\i}nez, L.},
  doi              = {10.1002/jcc.10216},
  journal          = {J. Comput. Chem.},
  pages            = {819--825},
  title            = {Packing optimization for automated generation of complex system{\textquoteright}s initial configurations for molecular dynamics and docking},
  volume           = {24},
  year             = {2003},
  creationdate     = {2025-08-20T15:46:39},
  issn             = {1096-987X},
  modificationdate = {2025-11-03T09:14:41},
  publisher        = {Wiley},
}

@Article{xTB_GFN-FF,
  author           = {Spicher, Sebastian and Grimme, Stefan},
  doi              = {10.1002/anie.202004239},
  journal          = {Angewandte Chemie International Edition},
  pages            = {15665--15673},
  title            = {Robust atomistic modeling of materials, organometallic, and biochemical systems},
  volume           = {59},
  year             = {2020},
  creationdate     = {2025-01-31T13:39:49},
  issn             = {1521-3773},
  modificationdate = {2025-11-28T15:01:32},
  publisher        = {Wiley},
}

@Book{Berendsen_2007,
  author           = {Berendsen, Herman J. C.},
  publisher        = {Cambridge University Press},
  title            = {Simulating the physical world: hierarchical modeling from quantum mechanics to fluid dynamics},
  year             = {2007},
  isbn             = {9780511815348},
  creationdate     = {2025-12-01T17:58:52},
  doi              = {10.1017/CBO9780511815348},
  modificationdate = {2025-12-01T17:59:43},
}

@Article{xTB_MTD,
  author           = {S. Grimme},
  doi              = {10.1021/acs.jctc.9b00143},
  journal          = {J. Chem. Theory Comput.},
  pages            = {2847--2862},
  title            = {Exploration of chemical compound, conformer, and reaction
		  space with meta-dynamics simulations based on tight-binding
		  quantum chemical calculations},
  volume           = {15},
  year             = {2019},
  creationdate     = {2023-02-28T15:52:17},
  modificationdate = {2023-02-28T15:56:11},
  publisher        = {American Chemical Society ({ACS})},
}

@Book{Hansen_2013,
  author           = {Jean-Pierre Hansen and I. R. McDonald},
  publisher        = {Academic Press},
  title            = {Theory of simple liquids: with applications to soft matter},
  year             = {2013},
  creationdate     = {2025-08-21T15:06:44},
  modificationdate = {2025-11-28T15:01:32},
}

@Article{Ribeiro2017,
  author           = {Ribeiro, M. S. and Pascoini, A. L. and Knupp, W. G. and Camps, I.},
  doi              = {10.1016/j.apsusc.2017.07.162},
  journal          = {Appl. Surf. Sci.},
  pages            = {781--787},
  title            = {Effects of surface functionalization on the electronic and structural properties of carbon nanotubes: A computational approach},
  volume           = {426},
  year             = {2017},
  creationdate     = {2026-08-12T10:08:37},
  issn             = {0169-4332},
  modificationdate = {2026-08-12T10:08:37},
  publisher        = {Elsevier BV},
}

@Article{Oganov2009,
  author           = {Artem R. Oganov and Mario Valle},
  journal          = {J. Chem. Phys.},
  pages            = {104504},
  title            = {How to quantify energy landscapes of solids},
  volume           = {130},
  year             = {2009},
  creationdate     = {2026-08-12T10:11:54},
  modificationdate = {2026-08-12T10:11:54},
}

@Article{xTB_ALPB,
  author           = {Ehlert, Sebastian and Stahn, Marcel and Spicher, Sebastian and Grimme, Stefan},
  doi              = {10.1021/acs.jctc.1c00471},
  journal          = {J. Chem. Theory Comput.},
  pages            = {4250--4261},
  title            = {Robust and efficient implicit solvation model for fast semiempirical methods},
  volume           = {17},
  year             = {2021},
  creationdate     = {2026-08-12T15:06:53},
  issn             = {1549-9626},
  modificationdate = {2026-08-12T15:07:09},
  publisher        = {American Chemical Society (ACS)},
}

@Article{Shekhar-ToxicolRep-13-101840-2024,
  author           = {Shekhar, Chander and Khosya, Reetu and Thakur, Kushal and Mahajan, Danish and Kumar, Rakesh and Kumar, Sunil and Sharma, Amit Kumar},
  doi              = {10.1016/j.toxrep.2024.101840},
  journal          = {Toxicol. Rep.},
  pages            = {101840},
  title            = {A systematic review of pesticide exposure, associated risks, and long-term human health impacts},
  volume           = {13},
  year             = {2024},
  creationdate     = {2026-08-12T15:48:09},
  modificationdate = {2026-08-12T15:48:09},
  publisher        = {Elsevier BV},
}

@Article{Munoz-Bautista-Agronomy-15-1878-2025,
  author           = {Mu{\~{n}}oz-Bautista, Jes{\'{u}}s Mart{\'{i}}n and Bernal-Mercado, Ariadna Thal{\'{i}}a and Mart{\'{i}}nez-Cruz, Oliviert and Burgos-Hern{\'{a}}ndez, Armando and L{\'{o}}pez-Zavala, Alonso Alexis and Ruiz-Cruz, Saul and Ornelas-Paz, Jos{\'{e}} de Jes{\'{u}}s and Borboa-Flores, Jes{\'{u}}s and Ramos-Enr{\'{i}}quez, Jos{\'{e}} Rogelio and Del-Toro-S{\'{a}}nchez, Carmen Lizette},
  doi              = {10.3390/agronomy15081878},
  journal          = {Agronomy},
  pages            = {1878},
  title            = {Environmental and health impacts of pesticides and nanotechnology as an alternative in agriculture},
  volume           = {15},
  year             = {2025},
  creationdate     = {2026-08-12T15:50:57},
  modificationdate = {2026-08-12T15:50:57},
  publisher        = {MDPI AG},
}

@Article{Zhou-EmergingContaminants-11-100410-2025,
  author           = {Zhou, Wei and Li, Mengmeng and Achal, Varenyam},
  doi              = {10.1016/j.emcon.2024.100410},
  journal          = {Emerging Contaminants},
  pages            = {100410},
  title            = {A comprehensive review on environmental and human health impacts of chemical pesticide usage},
  volume           = {11},
  year             = {2025},
  creationdate     = {2026-08-12T15:53:36},
  modificationdate = {2026-08-12T15:53:36},
  publisher        = {Elsevier BV},
}

@Article{Lazarevic-Pasti-Foods-14-1128-2025,
  author           = {Lazarevi{\'{c}}-Pa{\v{s}}ti, Tamara and Milankovi{\'{c}}, Vedran and Tasi{\'{c}}, Tamara and Petrovi{\'{c}}, Sandra and Leskovac, Andreja},
  doi              = {10.3390/foods14071128},
  journal          = {Foods},
  pages            = {1128},
  title            = {With or without you? - {A} critical review on pesticides in food},
  volume           = {14},
  year             = {2025},
  creationdate     = {2026-08-12T15:56:13},
  modificationdate = {2026-08-12T15:56:13},
  publisher        = {MDPI AG},
}

@Article{Galli-FrontiersinToxicology-6-1474792-2024,
  author           = {Galli, Flavia Silvia and Mollari, Marta and Tassinari, Valentina and Alimonti, Cristian and Ubaldi, Alessandro and Cuva, Camilla and Marcoccia, Daniele},
  doi              = {10.3389/ftox.2024.1474792},
  journal          = {Frontiers in Toxicology},
  pages            = {1474792},
  title            = {Overview of human health effects related to glyphosate exposure},
  volume           = {6},
  year             = {2024},
  creationdate     = {2026-08-12T15:58:20},
  modificationdate = {2026-08-12T15:58:20},
  publisher        = {Frontiers Media SA},
}

@Article{Badani-EurJEnvironSci-13-5-2023,
  author           = {Badani, Hadjer and Djadouni, Fatima and Haddad, Fatma Zohra},
  doi              = {10.14712/23361964.2023.1},
  journal          = {Eur. J. Environ. Sci.},
  pages            = {5--14},
  title            = {Effects of the herbicide glyphosate [n-(phosphonomethyl) glycine] on biodiversity and organisms in the soil},
  volume           = {13},
  year             = {2023},
  creationdate     = {2026-08-12T16:00:51},
  modificationdate = {2026-08-12T16:00:51},
  publisher        = {Charles University in Prague, Karolinum Press},
}

@Article{KimbiYaah-EnvironRes-240-117477-2024,
  author           = {Kimbi Yaah, Velma Beri and Ahmadi, Sajad and Quimbayo M, Jennyffer and Morales-Torres, Sergio and Ojala, Satu},
  doi              = {10.1016/j.envres.2023.117477},
  journal          = {Environ. Res.},
  pages            = {117477},
  title            = {Recent technologies for glyphosate removal from aqueous environment: A critical review},
  volume           = {240},
  year             = {2024},
  creationdate     = {2026-08-12T16:03:05},
  modificationdate = {2026-08-12T16:03:05},
  publisher        = {Elsevier BV},
}

@Article{Chavez-Reyes-JXenobiot-14-604-2024,
  author           = {Ch{\'{a}}vez-Reyes, Jes{\'{u}}s and Sar{\'{a}}chaga-Terrazas, Fernando and Colis-Arenas, Oliver Alejandro and L{\'{o}}pez-Lariz, Carlos H. and Villal{\'{o}}n, Carlos M. and Marichal-Cancino, Bruno A.},
  doi              = {10.3390/jox14020035},
  journal          = {J. Xenobiot.},
  pages            = {604--612},
  title            = {Aminomethylphosphonic acid ({AMPA}), a glyphosate metabolite, decreases plasma cholinesterase activity in rats},
  volume           = {14},
  year             = {2024},
  creationdate     = {2026-08-12T16:04:51},
  modificationdate = {2026-08-12T16:04:51},
  publisher        = {MDPI AG},
}

@Article{Yubolphan-EnvironToxicolPhar-125-105094-2026,
  author           = {Yubolphan, Ruedeemars and Kantisin, Siriwan and Satayavivad, Jutamaad and Essigmann, John M. and Fedeles, Bogdan I. and Konguthaithip, Giatgong and Tajai, Preechaya},
  doi              = {10.1016/J.ETAP.2026.105094},
  journal          = {Environ. Toxicol. Phar.},
  pages            = {105094},
  title            = {Glyphosate-based herbicides induce oxidative stress: Insights from human biomarker and in vitro mutagenic studies},
  volume           = {125},
  year             = {2026},
  creationdate     = {2026-08-12T16:06:49},
  modificationdate = {2026-08-12T16:06:49},
  publisher        = {Elsevier BV},
}

@Article{Kalyabina-ToxicolRep-8-1179-2021,
  author           = {Kalyabina, Valeriya P. and Esimbekova, Elena N. and Kopylova, Kseniya V. and Kratasyuk, Valentina A.},
  doi              = {10.1016/j.toxrep.2021.06.004},
  journal          = {Toxicol. Rep.},
  pages            = {1179--1192},
  title            = {Pesticides: formulants, distribution pathways and effects on human health – a review},
  volume           = {8},
  year             = {2021},
  creationdate     = {2026-08-12T16:07:58},
  issn             = {2214-7500},
  modificationdate = {2026-08-12T16:07:58},
  publisher        = {Elsevier BV},
}

@Article{Barroso-Sustainability-17-3891-2025,
  author           = {Barroso, Gabriela Madureira and Leite, Maehssa Leonor Franco and Silva, Gabriele Gon{\c{c}}alves and Barboza, Heliene Meira and Pinto, Thiago Almeida Andrade and da Costa, M{\'{a}}rcia Regina and Aguiar, Luciana Monteiro and da Silva Te{\'{o}}filo, Taliane Maria and dos Santos, Jos{\'{e}} Barbosa},
  doi              = {10.3390/su17093891},
  journal          = {Sustainability},
  pages            = {3891},
  title            = {Pesticide residue management in {Brazil}: implications for human health and the environment},
  volume           = {17},
  year             = {2025},
  creationdate     = {2026-08-12T16:09:18},
  modificationdate = {2026-08-12T16:09:18},
  publisher        = {MDPI AG},
}

@Article{Ray-EnvironAnalHealToxicol-38-2023017-2023,
  author           = {Ray, Suryapratap and Shaju, Sanjana Thanjan},
  doi              = {10.5620/eaht.2023017},
  journal          = {Environ. Anal. Heal. Toxicol.},
  pages            = {e2023017},
  title            = {Bioaccumulation of pesticides in fish resulting toxicities in humans through food chain and forensic aspects},
  volume           = {38},
  year             = {2023},
  creationdate     = {2026-08-12T16:12:04},
  modificationdate = {2026-08-12T16:12:04},
  publisher        = {The Korean Society of Environmental Health and Toxicology},
}

@Article{Sittiwong-MicroporMesoporMat-341-112083-2022,
  author           = {Sittiwong, Jarinya and Hiruntrakool, Keeradara and Rasrichai, Athittaya and Opasmongkolchai, Ornanong and Srifa, Pemika and Nilwanna, Krongkwan and Maihom, Thana and Probst, Michael and Limtrakul, Jumras},
  doi              = {10.1016/J.MICROMESO.2022.112083},
  journal          = {Micropor. Mesopor. Mat.},
  pages            = {112083},
  title            = {Insights into glyphosate adsorption on {Lewis} acidic zeolites from theoretical modelling},
  volume           = {341},
  year             = {2022},
  creationdate     = {2026-08-12T16:16:13},
  modificationdate = {2026-08-12T16:16:13},
  publisher        = {Elsevier BV},
}

@Article{Li-EnvironSciPollutR-25-21036-2018,
  author           = {Li, Yajuan and Zhao, Chuanqi and Wen, Yujuan and Wang, Yuanyuan and Yang, Yuesuo},
  doi              = {10.1007/s11356-018-2282-x},
  journal          = {Environ. Sci. Pollut. R.},
  pages            = {21036--21048},
  title            = {Adsorption performance and mechanism of magnetic reduced graphene oxide in glyphosate contaminated water},
  volume           = {25},
  year             = {2018},
  creationdate     = {2026-08-12T16:17:27},
  modificationdate = {2026-08-12T16:17:27},
  publisher        = {Springer Science and Business Media LLC},
}

@Article{Wang-Chemosphere-331-138827-2023,
  author           = {Wang, Qi and Cui, Kang-Ping and Liu, Tong and Li, Chen-Xuan and Liu, Jun and Kong, Dian-Chao and Weerasooriya, Rohan and Chen, Xing},
  doi              = {10.1016/J.CHEMOSPHERE.2023.138827},
  journal          = {Chemosphere},
  pages            = {138827},
  title            = {In situ growth of {NH2-MIL-101} metal organic frameworks on biochar for glyphosate adsorption},
  volume           = {331},
  year             = {2023},
  creationdate     = {2026-08-12T16:18:34},
  modificationdate = {2026-08-12T16:18:34},
  publisher        = {Elsevier BV},
}

@Article{Singh-IntJEnvResPubHe-17-7519-2020,
  author           = {Singh, Simranjeet and Kumar, Vijay and Gill, Jatinder Pal Kaur and Datta, Shivika and Singh, Satyender and Dhaka, Vaishali and Kapoor, Dhriti and Wani, Abdul Basit and Dhanjal, Daljeet Singh and Kumar, Manoj and Harikumar, S. L. and Singh, Joginder},
  doi              = {10.3390/ijerph17207519},
  journal          = {Int. J. Env. Res. Pub. He.},
  pages            = {7519},
  title            = {Herbicide glyphosate: toxicity and microbial degradation},
  volume           = {17},
  year             = {2020},
  creationdate     = {2026-08-12T16:20:02},
  modificationdate = {2026-08-12T16:20:02},
  publisher        = {MDPI AG},
}

@Article{SalgadoKiefer-EnvironToxicolPhar-107-104429-2024,
  author           = {Salgado Kiefer, Yvanna Carla de Souza and Ferreira, Marianna Boia and da Luz, Jessica Zablocki and Filipak Neto, Francisco and Oliveira Ribeiro, Ciro Alberto de},
  doi              = {10.1016/j.etap.2024.104429},
  journal          = {Environ. Toxicol. Phar.},
  pages            = {104429},
  title            = {Glyphosate and aminomethylphosphonic acid metabolite ({AMPA}) modulate the phenotype of murine melanoma {B16-F1} cells},
  volume           = {107},
  year             = {2024},
  creationdate     = {2026-08-12T16:21:36},
  issn             = {1382-6689},
  modificationdate = {2026-08-12T16:21:36},
  month            = Apr,
  publisher        = {Elsevier BV},
}

@Article{Grandcoin-WaterRes-117-187-2017,
  author           = {Grandcoin, Alexis and Piel, St{\'{e}}phanie and Baur{\`{e}}s, Estelle},
  doi              = {10.1016/J.WATRES.2017.03.055},
  journal          = {Water Res.},
  pages            = {187--197},
  title            = {{AminoMethylPhosphonic acid (AMPA}) in natural waters: Its sources, behavior and environmental fate},
  volume           = {117},
  year             = {2017},
  creationdate     = {2026-08-12T16:23:15},
  modificationdate = {2026-08-12T16:23:15},
  publisher        = {Elsevier BV},
}

@Article{Saleh-JIndEngChem-146-176-2025,
  author           = {Saleh, Mohammed and Gul, Afroz and Nasir, Abir and Moses, Titus Otamayomi and Nural, Yahya and Yabalak, Erdal},
  doi              = {10.1016/J.JIEC.2024.11.052},
  journal          = {J. Ind. Eng. Chem.},
  pages            = {176--212},
  title            = {Comprehensive review of carbon-based nanostructures: {P}roperties, synthesis, characterization, and cross-disciplinary applications},
  volume           = {146},
  year             = {2025},
  creationdate     = {2026-08-13T15:19:07},
  modificationdate = {2026-08-13T15:19:07},
  publisher        = {Elsevier BV},
}

@Article{Sandoval-ChemRev-126-2283-2026,
  author           = {Sandoval, Stefania and Gon{\c{c}}alves, Gil and P{\'{e}}rez Barrio, Jorge and Kharlamova, Marianna V. and Tob{\'{i}}as-Rossell, Gerard},
  doi              = {10.1021/acs.chemrev.5c00219},
  journal          = {Chem. Rev.},
  pages            = {2283--2390},
  title            = {A comprehensive review on filled carbon nanotubes: {S}ynthesis, properties and applications},
  volume           = {126},
  year             = {2026},
  creationdate     = {2026-08-13T15:20:29},
  modificationdate = {2026-08-13T15:20:29},
  publisher        = {American Chemical Society (ACS)},
}

@Article{Alfei-JXenobiotics-15-76-2025,
  author           = {Alfei, Silvana and Zuccari, Guendalina},
  doi              = {10.3390/jox15030076},
  journal          = {J. Xenobiotics},
  pages            = {76},
  title            = {Carbon-nanotube-based nanocomposites in environmental remediation: {A}n overview of typologies and applications and an analysis of their paradoxical double-sided effects},
  volume           = {15},
  year             = {2025},
  creationdate     = {2026-08-13T15:22:17},
  modificationdate = {2026-08-13T15:22:17},
  publisher        = {MDPI AG},
}

@Article{Alosta-MaterTodayCommun-50-114504-2026,
  author           = {Alosta, Ebrahim and Kumari, Priyanka and Rashed, Ahmed and Aubry, Cyril and Low, Nicholas (Ze-Xian) and Huynh, Chi and Dum{\'{e}}e, Ludovic F.},
  doi              = {10.1016/j.mtcomm.2025.114504},
  journal          = {Mater. Today Commun.},
  pages            = {114504},
  title            = {Hydrophilic carbon nanotube-based electrodes for enhanced hydrogen evolution reaction by defect engineering control},
  volume           = {50},
  year             = {2026},
  creationdate     = {2026-08-13T15:23:12},
  modificationdate = {2026-08-13T15:23:12},
  publisher        = {Elsevier BV},
}

@Article{Dong-ChemEngJ-499-156654-2024,
  author           = {Dong, Wei and Xing, Jing and Chen, Quan and Huang, Yu and Wu, Min and Yi, Peng and Pan, Bo and Xing, Baoshan},
  doi              = {10.1016/J.CEJ.2024.156654},
  journal          = {Chem. Eng. J.},
  pages            = {156654},
  title            = {Hydrogen bonds between the oxygen-containing functional groups of biochar and organic contaminants significantly enhance sorption affinity},
  volume           = {499},
  year             = {2024},
  creationdate     = {2026-08-13T15:25:11},
  modificationdate = {2026-08-13T15:25:11},
  publisher        = {Elsevier BV},
}

@Article{Milowska-JChemPhys-138-194704-2013,
  author           = {Milowska, Karolina Z. and Majewski, Jacek A.},
  doi              = {10.1063/1.4804652},
  journal          = {J. Chem. Phys.},
  pages            = {194704},
  title            = {Functionalization of carbon nanotubes with {-CHn}, {-NHn} fragments, {-COOH and -OH} groups},
  volume           = {138},
  year             = {2013},
  creationdate     = {2026-08-13T15:27:34},
  modificationdate = {2026-08-13T15:27:34},
  publisher        = {AIP Publishing},
}

@Article{Rezazade-BMCChemistry-18-85-2024,
  author           = {Rezazade, Masume and Ketabi, Sepideh and Qomi, Mahnaz},
  doi              = {10.1186/s13065-024-01197-0},
  journal          = {BMC Chemistry},
  pages            = {85},
  title            = {Effect of functionalization on the adsorption performance of carbon nanotube as a drug delivery system for imatinib: molecular simulation study},
  volume           = {18},
  year             = {2024},
  creationdate     = {2026-08-13T15:29:17},
  modificationdate = {2026-08-13T15:29:17},
  publisher        = {Springer Science and Business Media LLC},
}

@Article{Silva-ApplSurfSci-729-166060-2026,
  author           = {Silva, H. T. and Faria, L. C. S. and Aversi-Ferreira, T. A. and Camps, I.},
  doi              = {10.1016/j.apsusc.2026.166060},
  journal          = {Appl. Surf. Sci.},
  pages            = {166060},
  title            = {{pH-Responsive} glyphosate adsorption on hydroxylated carbon nanotubes: {From} electronic structure to molecular dynamics},
  volume           = {729},
  year             = {2026},
  creationdate     = {2026-08-13T15:30:59},
  issn             = {0169-4332},
  modificationdate = {2026-08-13T15:30:59},
  month            = May,
  publisher        = {Elsevier BV},
}

@Article{Mananghaya-JMolLiq-212-592-2015,
  author           = {Mananghaya, Michael},
  doi              = {10.1016/J.MOLLIQ.2015.10.013},
  journal          = {J. Mol. Liq.},
  pages            = {592--596},
  title            = {Modeling of single-walled carbon nanotubes functionalized with carboxylic and amide groups towards its solubilization in water},
  volume           = {212},
  year             = {2015},
  creationdate     = {2026-08-13T15:32:09},
  modificationdate = {2026-08-13T15:32:09},
  publisher        = {Elsevier BV},
}

@Article{Lara-ChemPhys-428-117-2014,
  author           = {Lara, Ivi Valentini and Zanella, Ivana and Fagan, Solange Binotto},
  doi              = {10.1016/J.CHEMPHYS.2013.11.007},
  journal          = {Chem. Phys.},
  pages            = {117--120},
  title            = {Functionalization of carbon nanotube by carboxyl group under radial deformation},
  volume           = {428},
  year             = {2014},
  creationdate     = {2026-08-13T15:36:10},
  modificationdate = {2026-08-13T15:36:10},
  publisher        = {Elsevier BV},
}

@Article{Veloso-ChemPhysLett-430-71-2006,
  author           = {Veloso, Marcos V. and Souza Filho, A. G. and Mendes Filho, J. and Fagan, Solange B. and Mota, R.},
  doi              = {10.1016/J.CPLETT.2006.08.082},
  journal          = {Chem. Phys. Lett.},
  pages            = {71--74},
  title            = {Ab initio study of covalently functionalized carbon nanotubes},
  volume           = {430},
  year             = {2006},
  creationdate     = {2026-08-13T15:37:13},
  modificationdate = {2026-08-13T15:37:13},
  publisher        = {Elsevier BV},
}

@Article{Silva-SurfacesandInterfaces-93-109439-2026,
  author           = {Silva, H. T. and Faria, L. C. S. and Aversi-Ferreira, T. A. and Camps, I.},
  doi              = {10.1016/j.surfin.2026.109439},
  journal          = {Surfaces and Interfaces},
  pages            = {109439},
  title            = {Computational study of interactions between ionized glyphosate and carbon nanotube: {An} alternative for mitigating environmental contamination},
  volume           = {93},
  year             = {2026},
  creationdate     = {2026-08-13T15:38:53},
  modificationdate = {2026-08-13T15:38:53},
  publisher        = {Elsevier BV},
}

@Article{Bulla-JEnvironChemEng-12-114504-2024,
  author           = {Bulla, Mamta and Kumar, Vinay and Devi, Raman and Kumar, Sunil and Dahiya, Rita and Singh, Parul and Mishra, Ajay Kumar},
  doi              = {10.1016/J.JECE.2024.114504},
  journal          = {J. Environ. Chem. Eng.},
  pages            = {114504},
  title            = {Exploring the frontiers of carbon nanotube synthesis techniques and their potential applications in supercapacitors, gas sensing, and water purification},
  volume           = {12},
  year             = {2024},
  creationdate     = {2026-08-13T15:39:47},
  modificationdate = {2026-08-13T15:39:47},
  publisher        = {Elsevier BV},
}

@Article{Cui-Nanomaterials-13-2781-2023,
  author           = {Cui, Hongyuan and Gao, Chenshan and Wang, Pengwei and Li, Lijie and Ye, Huaiyu and Wen, Zhongquan and Liu, Yufei},
  doi              = {10.3390/nano13202781},
  journal          = {Nanomaterials},
  pages            = {2781},
  title            = {{DFT} study of {Zn}-modified {SnP3: A H2S} gas sensor with superior sensitivity, selectivity, and fast recovery time},
  volume           = {13},
  year             = {2023},
  creationdate     = {2026-08-14T09:24:54},
  issn             = {2079-4991},
  modificationdate = {2026-08-14T09:24:54},
  publisher        = {MDPI AG},
}

@Article{Rahimi-SciRep-14-29282-2024,
  author           = {Rahimi, Rezvan and Solimannejad, Mohammad and Horri, Ashkan},
  doi              = {10.1038/s41598-024-77659-1},
  journal          = {Sci. Rep.},
  pages            = {29282},
  title            = {{DFT} study of the adsorption properties and sensitivity of a {B2N} monolayer toward harmful gases},
  volume           = {14},
  year             = {2024},
  creationdate     = {2026-08-14T09:25:43},
  issn             = {2045-2322},
  modificationdate = {2026-08-14T09:25:43},
  publisher        = {Springer Science and Business Media LLC},
}

@Misc{jmol,
  title            = {{Jmol: An open-source Java viewer for chemical structures in {3D}. http://www.jmol.org/}},
  creationdate     = {2026-08-19T14:46:17},
  modificationdate = {2026-08-19T14:46:43},
}

@Article{Bader-ChemRev-91-893-1991,
  author           = {Bader, Richard F. W.},
  doi              = {10.1021/cr00005a013},
  journal          = {Chem. Rev.},
  pages            = {893--928},
  title            = {A quantum theory of molecular structure and its applications},
  volume           = {91},
  year             = {1991},
  creationdate     = {2026-08-28T10:24:41},
  modificationdate = {2026-08-28T10:24:41},
  publisher        = {American Chemical Society (ACS)},
}

@Article{Shahbazian-ChemEurJ-24-5401-2018,
  author           = {Shahbazian, Shant},
  doi              = {10.1002/chem.201705163},
  journal          = {Chem. Eur. J.},
  pages            = {5401--5405},
  title            = {Why bond critical points are not ``bond'' critical points},
  volume           = {24},
  year             = {2018},
  creationdate     = {2026-08-28T10:29:05},
  modificationdate = {2026-08-28T10:29:05},
  publisher        = {Wiley},
}

@Article{Popelier-JMolModel-28-276-2022,
  author           = {Popelier, Paul L. A.},
  doi              = {10.1007/s00894-022-05188-7},
  journal          = {J. Mol. Model.},
  pages            = {276},
  title            = {Non-covalent interactions from a quantum chemical topology perspective},
  volume           = {28},
  year             = {2022},
  creationdate     = {2026-08-28T10:41:46},
  modificationdate = {2026-08-28T10:41:46},
  publisher        = {Springer Science and Business Media LLC},
}

@Article{Becke-JChemPhys-92-5397-1990,
  author           = {A. D. Becke and K. E. Edgecombe},
  doi              = {10.1063/1.458517},
  journal          = {J. Chem. Phys.},
  pages            = {5397--5403},
  title            = {A simple measure of electron localization in atomic and molecular systems},
  volume           = {92},
  year             = {1990},
  creationdate     = {2026-08-28T10:44:42},
  modificationdate = {2026-08-28T10:44:42},
  publisher        = {{AIP} Publishing},
}

@Article{Schmider-JMolStructTHEOCHEM-527-51-2000,
  author           = {Schmider, H. L. and Becke, A. D.},
  doi              = {10.1016/S0166-1280(00)00477-2},
  journal          = {J. Mol. Struct.: {THEOCHEM}},
  pages            = {51--61},
  title            = {Chemical content of the kinetic energy density},
  volume           = {527},
  year             = {2000},
  creationdate     = {2026-08-28T10:49:04},
  modificationdate = {2026-08-28T10:49:13},
  publisher        = {Elsevier {BV}},
}

@Article{Lu-JComputChem-33-580-2012,
  author           = {Lu, Tian and Chen, Feiwu},
  doi              = {10.1002/jcc.22885},
  journal          = {J. Comput. Chem.},
  pages            = {580--592},
  title            = {Multiwfn: A multifunctional wavefunction analyzer},
  volume           = {33},
  year             = {2012},
  creationdate     = {2026-08-28T10:53:58},
  issn             = {1096-987X},
  modificationdate = {2026-08-28T10:53:58},
  publisher        = {Wiley},
}

@Article{Rozas-JAmChemSoc-122-11154-2000,
  author           = {Rozas, Isabel and Alkorta, Ibon and Elguero, Jos{\'{e}}},
  doi              = {10.1021/ja0017864},
  journal          = {J. Am. Chem. Soc.},
  pages            = {11154--11161},
  title            = {Behavior of ylides containing {N, O, and C} atoms as hydrogen bond acceptors},
  volume           = {122},
  year             = {2000},
  creationdate     = {2026-08-28T10:58:05},
  modificationdate = {2026-08-28T10:58:05},
  publisher        = {American Chemical Society (ACS)},
}

@Article{Grabowski-ChemRev-111-2597-2011,
  author           = {Grabowski, S{{\l}}awomir Janusz},
  doi              = {10.1021/cr800346f},
  journal          = {Chem. Rev.},
  pages            = {2597--2625},
  title            = {What is the covalency of hydrogen bonding?},
  volume           = {111},
  year             = {2011},
  creationdate     = {2026-08-28T11:00:41},
  modificationdate = {2026-08-28T11:00:41},
  publisher        = {American Chemical Society (ACS)},
}

@Article{Espinosa-ChemPhysLett-285-170-1998,
  author           = {Espinosa, E. and Molins, E. and Lecomte, C.},
  doi              = {10.1016/S0009-2614(98)00036-0},
  journal          = {Chem. Phys. Lett.},
  pages            = {170--173},
  title            = {Hydrogen bond strengths revealed by topological analyses of experimentally observed electron densities},
  volume           = {285},
  year             = {1998},
  creationdate     = {2026-08-28T11:07:03},
  modificationdate = {2026-08-28T11:07:03},
  publisher        = {Elsevier BV},
}

\newpage

\pagebreak \clearpage

\begin{center}
	\textbf{Supplemental Materials} \\
	\textbf{From Electronic Structure to Environmental Remediation: Adsorption of Ionized Glyphosate on COOH--Modified Carbon Nanotubes}
\end{center}

\setcounter{section}{0}
\setcounter{equation}{0} \setcounter{figure}{0} \setcounter{table}{0}
\setcounter{page}{1} \makeatletter
\renewcommand{\thesection}{S\arabic{section}}
\renewcommand{\theequation}{S\arabic{equation}}
\renewcommand{\thefigure}{S\arabic{figure}}
\renewcommand{\thetable}{S\arabic{table}}

\tiny{
\begin{table}[htbp]
	\centering
	\setlength{\tabcolsep}{4pt}
	\caption{Topological descriptors$^\dagger$.}
	\label{STab:ElectProp}
	\renewcommand{\arraystretch}{0.85}
	\setlength\extrarowheight{-3pt}
	\begin{center}
	\resizebox{12cm}{!}{
	\begin{tabular}{lcccccccccc}
		\hline
		System        & BCP & Connected atoms  & $\rho(r)$ & $\nabla^2\rho(r)$ & G(r)     & V(r)      & H(r)      & |V(r)|/G(r) & ELF   & LOL   \\ \hline
		CNT+G5        & 304 & H130$\cdots$H152 & 0.004  & 0.020           & 0.003 & -0.002 & 0.002  & 0.555 & 0.009 & 0.085 \\
		CNT+G5        & 364 & H121$\cdots$P141 & 0.011  & 0.046           & 0.009 & -0.007 & 0.002  & 0.772 & 0.024 & 0.135 \\
		CNT+G5        & 375 & H121$\cdots$H155 & 0.007  & 0.035           & 0.007 & -0.004 & 0.002  & 0.659 & 0.014 & 0.106 \\
		CNT+G5        & 382 & H121$\cdots$O142 & 0.011  & 0.058           & 0.011 & -0.007 & 0.004  & 0.679 & 0.018 & 0.118 \\
		CNT+G5        & 407 & O144$\cdots$H123 & 0.015  & 0.063           & 0.013 & -0.010 & 0.003  & 0.803 & 0.035 & 0.159 \\
		CNT+G5        & 415 & O142$\cdots$C8   & 0.249  & -0.355          & 0.158 & -0.440 & -0.281 & 2.778 & 0.759 & 0.639 \\
		CNT+COOH5+G5  & 448 & O141$\cdots$O172 & 0.015  & 0.093           & 0.018 & -0.013 & 0.005  & 0.722 & 0.022 & 0.131 \\
		CNT+COOH5+G5  & 450 & C139$\cdots$O171 & 0.240  & -0.288          & 0.155 & -0.401 & -0.246 & 2.592 & 0.746 & 0.631 \\
		CNT+COOH10+G5 & 61  & O200$\cdots$C27  & 0.021  & 0.068           & 0.017 & -0.016 & 0.000  & 0.979 & 0.066 & 0.211 \\
		CNT+COOH10+G5 & 75  & O201$\cdots$O126 & 0.024  & 0.115           & 0.026 & -0.024 & 0.003  & 0.902 & 0.048 & 0.183 \\
		CNT+COOH10+G5 & 210 & O198$\cdots$C139 & 0.011  & 0.062           & 0.012 & -0.008 & 0.004  & 0.705 & 0.017 & 0.115 \\
		CNT+COOH10+G5 & 214 & C139$\cdots$O199 & 0.012  & 0.072           & 0.014 & -0.009 & 0.004  & 0.687 & 0.017 & 0.116 \\
		CNT+COOH10+G5 & 231 & O199$\cdots$C136 & 0.014  & 0.096           & 0.018 & -0.012 & 0.006  & 0.681 & 0.017 & 0.117 \\
		CNT+COOH10+G5 & 245 & H207$\cdots$O138 & 0.006  & 0.040           & 0.007 & -0.004 & 0.003  & 0.575 & 0.006 & 0.071 \\
		CNT+COOH10+G5 & 303 & O200$\cdots$H180 & 0.006  & 0.043           & 0.008 & -0.005 & 0.003  & 0.614 & 0.007 & 0.076 \\
		CNT+COOH10+G5 & 320 & O201$\cdots$H180 & 0.007  & 0.047           & 0.009 & -0.005 & 0.003  & 0.612 & 0.008 & 0.081 \\
		CNT+COOH15+G4 & 8   & O229$\cdots$O174 & 0.021  & 0.104           & 0.023 & -0.020 & 0.003  & 0.881 & 0.037 & 0.163 \\
		CNT+COOH15+G4 & 92  & O228$\cdots$C64  & 0.017  & 0.063           & 0.014 & -0.013 & 0.001  & 0.910 & 0.048 & 0.183 \\
		CNT+COOH15+G4 & 276 & O230$\cdots$C67  & 0.006  & 0.050           & 0.009 & -0.005 & 0.004  & 0.582 & 0.005 & 0.063 \\
		CNT+COOH15+G4 & 305 & H235$\cdots$C65  & 0.002  & 0.013           & 0.002 & -0.001 & 0.001  & 0.485 & 0.003 & 0.051 \\
		CNT+COOH15+G4 & 369 & O228$\cdots$C157 & 0.004  & 0.029           & 0.005 & -0.003 & 0.002  & 0.542 & 0.002 & 0.046 \\
		CNT+COOH15+G4 & 380 & O227$\cdots$C23  & 0.004  & 0.026           & 0.004 & -0.002 & 0.002  & 0.547 & 0.004 & 0.057 \\
		CNT+COOH20+G4 & 205 & O185$\cdots$O255 & 0.033  & 0.142           & 0.033 & -0.031 & 0.002  & 0.928 & 0.077 & 0.224 \\
		CNT+COOH20+G4 & 575 & C90$\cdots$O255  & 0.003  & 0.026           & 0.004 & -0.002 & 0.002  & 0.530 & 0.002 & 0.047 \\
		CNT+COOH20+G4 & 620 & C89$\cdots$H263  & 0.001  & 0.006           & 0.001 & -0.000 & 0.001  & 0.479 & 0.001 & 0.032 \\
		CNT+COOH20+G4 & 647 & O132$\cdots$O256 & 0.031  & 0.135           & 0.031 & -0.028 & 0.003  & 0.915 & 0.074 & 0.221 \\
		CNT+COOH20+G4 & 648 & O131$\cdots$P253 & 0.012  & 0.060           & 0.012 & -0.008 & 0.003  & 0.718 & 0.023 & 0.134 \\
		CNT+COOH20+G4 & 657 & O192$\cdots$H263 & 0.003  & 0.020           & 0.003 & -0.002 & 0.002  & 0.525 & 0.002 & 0.044 \\
		CNT+COOH20+G4 & 658 & O131$\cdots$H267 & 0.006  & 0.041           & 0.007 & -0.004 & 0.003  & 0.586 & 0.006 & 0.069 \\
		CNT+COOH20+G4 & 671 & O192$\cdots$N259 & 0.004  & 0.034           & 0.006 & -0.003 & 0.003  & 0.541 & 0.002 & 0.046 \\
		CNT+COOH20+G4 & 674 & C260$\cdots$O189 & 0.004  & 0.029           & 0.005 & -0.003 & 0.002  & 0.541 & 0.003 & 0.053 \\
		CNT+COOH20+G4 & 682 & O189$\cdots$O258 & 0.004  & 0.032           & 0.005 & -0.003 & 0.002  & 0.545 & 0.003 & 0.054 \\
		CNT+COOH25+G5 & 23  & O284$\cdots$O198 & 0.019  & 0.099           & 0.021 & -0.018 & 0.003  & 0.844 & 0.032 & 0.155 \\
		CNT+COOH25+G5 & 283 & O285$\cdots$O215 & 0.030  & 0.129           & 0.030 & -0.028 & 0.002  & 0.919 & 0.075 & 0.222 \\
		CNT+COOH25+G5 & 287 & H294$\cdots$O215 & 0.004  & 0.027           & 0.005 & -0.003 & 0.002  & 0.543 & 0.004 & 0.059 \\
		CNT+COOH25+G5 & 290 & H294$\cdots$O219 & 0.005  & 0.035           & 0.006 & -0.004 & 0.003  & 0.591 & 0.005 & 0.068 \\
		CNT+COOH25+G5 & 293 & H293$\cdots$O216 & 0.005  & 0.033           & 0.006 & -0.003 & 0.003  & 0.542 & 0.004 & 0.060 \\
		CNT+COOH25+G5 & 294 & H294$\cdots$O216 & 0.005  & 0.033           & 0.006 & -0.003 & 0.003  & 0.545 & 0.004 & 0.060 \\
		CNT+COOH25+G5 & 303 & H291$\cdots$O200 & 0.003  & 0.017           & 0.003 & -0.001 & 0.001  & 0.512 & 0.003 & 0.052 \\
		CNT+COOH25+G5 & 311 & O284$\cdots$O197 & 0.058  & 0.214           & 0.054 & -0.055 & -0.001 & 1.016 & 0.178 & 0.318 \\
		CNT+COOH25+G5 & 320 & O283$\cdots$O201 & 0.006  & 0.037           & 0.007 & -0.004 & 0.003  & 0.613 & 0.005 & 0.069 \\
		CNT+COOH25+G5 & 325 & O283$\cdots$O153 & 0.002  & 0.017           & 0.003 & -0.001 & 0.001  & 0.511 & 0.002 & 0.040
	\end{tabular}	
}
\begin{flushleft}
	\tiny {$^\dagger$ When required, units are atomic units.}
\end{flushleft}
\end{center}
\end{table}

\end{document}